\documentclass[aps,prd,superscriptaddress,nofootinbib,amsmath,amsfonts,preprintnumbers,notitlepage,10pt,english]{revtex4-1}
\usepackage{amsmath}
\usepackage{amssymb}
\usepackage{babel}

\makeatletter

\usepackage{array,multirow,graphicx}
\usepackage{dcolumn}
\usepackage{newlfont}
\usepackage{bm}
\usepackage[colorlinks,citecolor=blue,urlcolor=blue,linkcolor=blue]{hyperref}
\usepackage[figtopcap]{subfigure}
\usepackage{color}

\begin{document}

\date{\today}

\title{Anisotropic compact stars in higher--order curvature theory}

\author{G.G.L. Nashed}
\email{nashed@bue.edu.eg}
\affiliation {Centre for Theoretical Physics, The British University, P.O. Box
43, El Sherouk City, Cairo 11837, Egypt}

\author{S.D. Odintsov}
\email{odintsov@ieec.uab.es}
\affiliation{Institut de Ci\`encies de l'Espai (ICE-CSIC/IEEC),\\
Campus, c. Can Magrans s/n, 08193, Barcelona, Spain}
\affiliation{Instituci\'o Catalana de Recerca i Estudis Avan\c{c}ats (ICREA),
Barcelona, Spain}

\author{V.K. Oikonomou}
\email{v.k.oikonomou1979@gmail.com}
\affiliation{Department of Physics, Aristotle University of Thessaloniki, Thessaloniki 54124, Greece}
\affiliation{Laboratory for Theoretical Cosmology, Tomsk State University of Control Systems and Radioelectronics (TUSUR), 634050 Tomsk, Russia}

\begin{abstract}
In this paper we shall consider spherically symmetric spacetime
solutions describing the interior of stellar compact objects, in
the context of higher--order curvature theory of the
$\mathrm{f(R)}$ type. We shall derive the non--vacuum field
equations of the higher--order curvature theory, without assuming
any specific form of the $\mathrm{f(R)}$ theory, specifying the
analysis for a spherically symmetric spacetime with two unknown
functions. We obtain a system of highly non-linear differential
equations, which consists of four differential equations with six
unknown functions. To solve such a system, we assume a specific
form of metric potentials,  using the Krori-Barua ansatz. We
successfully solve the system of differential equations, and we
derive all the components of the energy--momentum tensor.
Moreover, we derive the non-trivial general form of
$\mathrm{f(R)}$ that may generate such solutions and calculate the
dynamic Ricci scalar of the anisotropic star. Accordingly, we
calculate the asymptotic form of the function $\mathrm{f(R)}$,
which is a polynomial function. We match  the derived interior
solution with the exterior one, which was derived in
\cite{Nashed:2019tuk}, with the latter also resulting to a
non-trivial form of the Ricci scalar. Notably but rather expected,
the exterior solution differs from the Schwarzschild one in the
context of general relativity. The matching procedure will
eventually relate two constants with the mass and radius of the
compact stellar object. We list the necessary conditions that any
compact anisotropic star must satisfy and explain in detail that
our model bypasses all of these conditions for  a special compact
star $\textit {Her X--1 }$, which has an estimated mass and radius
\textit {(mass = 0.85 $\pm 0.15M_{\circledcirc}$\,\, and\,
\,radius $= 8.1 \pm 0.41$km)}. Moreover, we study the stability of
this model by using the Tolman-Oppenheimer-Volkoff equation and
adiabatic index, and we show that the considered model is
different and  more stable compared to the corresponding models in
the context of general relativity.
\end{abstract}

\pacs{04.50.Kd, 04.25.Nx, 04.40.Nr}
\keywords{$\mathbf{F(R)}$ gravitational theory, analytic spherically symmetric black holes, thermodynamics, stability, geodesic deviation.}

\maketitle
\section{\bf Introduction}

Apart from the great successes of Newtonian gravity, it utterly
failed in certain cases where strong gravitational effects were
considered, such as the advances of Mercury in addition to the
Mickelson Morley experiment \cite{PhysRevLett.103.090401}. In
1915, Einstein developed the general theory of relativity (GR),
which enabled the resolution of the issue with Mercury
\cite{Wheeler:1990nd}. Thereafter, GR is considered as the
cornerstone theory for gravitational physics. However GR has
several shortcomings that indicate GR not being the most
fundamental theory of gravity, such as the dark energy issues
\cite{Perlmutter:1998np,Riess:1998cb,Riess:2004nr,Hirata:1987hu,Dodelson:1993je,Cole:1994ab}.
In addition, the GR  violates the Chandrasekhar mass-limit for
white dwarfs of super-Chandrasekhar, and sub-Chandrasekhar
limiting mass
\cite{Howell:2006vn,Scalzo:2010xd,Filippenko:1992wda,10.1093/mnras/284.1.151,Turatto:1998eq,2001PASP..113..308M,Garnavich:2001vx,Taubenberger:2007dt}.

Moreover, GR shows inconsistency in the regime of strong
gravitational field and recent observations
\cite{Hawkins:2002sg,Spergel:2006hy,Perlmutter:1998np,Shekh:2019msk}.
Thus seeking for an appropriate modification of GR, is a well
motivated task. The most successful modification of GR is the
higher--order--curvature theory, and specifically $\mathrm{f(R)}$
gravity, which is successful in  explaining the presence of dark
matter and confronting  gravitational theories with observations
\cite{Bamba:2012cp}. Moreover, the
$\mathrm{f(R)}$ gravitational theory when quantized results
 to a renormalizable gravitational theory \cite{Stelle:1976gc}. Thus,
$\mathrm{f(R)}$ gravitational theory certainly is an appealing and
well-motivated extension of GR. Modified gravity theories are
divided into different categories such as those containing some
four second--order curvature invariants and other that involve the
invariants as a function of the  Ricci scalar--like
$\mathrm{f(R)}$ gravity model
\cite{Vainio:2016qas,Capozziello:2018ddp,Nojiri:2007as,Tang:2019qiy,Awad2017,Nojiri:2007cq,Nashed2003841,Song:2007da,Awad2018,Nojiri:2006gh,ElHanafy20161,Li:2007xn,Zhang:2005vt,Pogosian:2007sw,Shirafuji19971355,PhysRevD.81.049901,Cognola:2007zu}. The  $\mathrm{f(R)}$  gravitational theory
avoids the Ostrogradsky's instability \cite{Ostrogradsky:1850fid}
which is a common  limitation of general higher--derivative
theories \cite{Woodard:2006nt}.

Numerous applications of $\mathrm{f(R)}$ can be found in the
context of theoretical cosmology
\cite{Nojiri:2019lqw,Vasilev:2019iuh,Nashed20062241,Shah:2019mxn,Oikonomou:2018npe,Nashed2010,Awad2018,Battye:2017ysh}
and in astrophysics. Spherically symmetric vacuum black hole
solutions in  $\mathrm{f(R)}$  have been  derived in
\cite{PhysRevD.74.064022,2018EPJP..133...18N,2018IJMPD..2750074N,Nashed:2018piz,
Capozziello_2007,Nashed2007851,2012GReGr..44.1881C,Capozziello_2010,Elizalde:2020icc,Nashed:2019yto,Nashed:2019tuk}.
In the frame of a strong gravitational background in local
objects, numerous  spherically symmetric  black holes are derived
\cite{Sultana:2018fkw,Canate:2017bao,Yu:2017uyd,Canate:2015dda,Nashed2008291,Kehagias:2015ata,PhysRevD.82.104026,delaCruzDombriz:2009et}.
Recently, the study of compact stars in amended gravitational
theories has become popular. Compact stars result from the
collapse of massive stars and there are several types of compact
objects of interest, including white dwarfs, neutron stars,
strange stars and black holes. Various models describe neutron
stars in $\mathrm{f(R)}$
\cite{Feng:2017hje,Astashenok:2020cqq,Resco:2016upv,Nashed:2020kjh,Capozziello:2015yza,Astashenok:2020cfv,Yazadjiev:2014cza,Nashed:2021sji,Ganguly:2013taa,Astashenok:2013vza,Orellana:2013gn,Arapoglu:2010rz,Cooney:2009rr}. Moreover hypernuclear compact stars is studied for stellar models constructed on the basis of covariant density functional theory in Hartree and HartreeFock approximation \cite{Raduta:2019rsk}.
In the present work we aim to  apply the non-vacuum field
equations of $\mathrm{f(R)}$ to a spherically symmetric spacetime
without assuming any specific form of $\mathrm{f(R)}$, and to
derive a compact anisotropic model. The resulting model shall be
confronted with real compact anisotropic stars,  and specifically
the star $\textit {Her X--1 }$.

The article is organized as follows: In Sec. \ref{S2}, we give a
brief summary of the $\mathrm{f(R)}$ gravitational theory. In Sec.
\ref{S3}, we apply the non-vacuum field equations of
$\mathrm{f(R)}$ to a spherically symmetric line-element that has
an unequal metric potential. We derive a system of differential
equations, having six unknown functions. In order to derive  an
analytic solution for the differential equations in closed form,
we assume a specific form of the metric potential, using the
Krori-Barua ansatz. We derive the remaining unknown functions, all
the components of the energy-momentum tensor, and the asymptotic
form of  the polynomial $\mathrm{f(R)}$ which generates such a
solution. This solution is characterized by four constants of
integration, and one of them differentiates our model from the
corresponding GR description. In Sec. \ref{S4}, we match the model
derived in Sec. \ref{S3},  with the exterior solution presented in
\cite{Nashed:2019tuk}, which has a spherically symmetric solution
different from the Schwarzschild one,  and successfully  match two
constants with the mass and radius of the compact stellar object.
In Sec. \ref{S5}, we list the necessary conditions that any
realistic theoretical model  must satisfy in order for it to
become compatible with a realistic star. We show that our model
satisfies all of these conditions that are required for any
realistic compact stellar object. In Sec. \ref{S6}, we study the
stability using the Tolman-Oppenheimer-Volkoff (TOV) equation and
adiabatic index and show that the present model satisfies these
requirements implying its stability. In the final section, we
present our concluding remarks.

\section{\bf Summary of the $\mathrm{f(R)}$ gravitational theory}\label{S2}

In this section, we consider recall the essential features of
four-dimensional higher--order curvature $\mathrm{f(R)}$ gravity.
$\mathrm{f(R)}$ gravity serves as a modification GR and coincides
with it when $\mathrm{f(R)=R}$. When $\mathrm{f(R)\neq
\mathrm{R}}$, we have a theory  different from Einstein's GR. The
action  of $\mathrm{f(R)}$ gravity  can take the following form
(cf.
\cite{Capozziello:2002rd,Carroll:2003wy,1970MNRAS.150....1B,Nojiri:2003ft,Capozziello:2003gx,Capozziello:2011et,Nojiri:2010wj,Nojiri:2017ncd}):
\begin{eqnarray} \label{a2} {\mathop{\mathcal{ I}}}:=\frac{1}{2\kappa} \int d^4x \sqrt{-g}  \mathrm{f(R)}+\frac{1}{2\kappa} \int d^4x \sqrt{-g} {\cal L_M}(g_{\mu \nu},\xi)\,,\end{eqnarray}
where $\kappa=8\pi G$, $G$ is Newton's  gravitational constant,
$g$ is the determinant of the metric, $ {\cal L_M}(g_{\mu
\nu},\xi)$ is the action of matter fields, and $\xi$ is minimally
coupled to the metric $g_{\mu \nu}$.

Upon varying the gravitational action  with respect to the metric
tensor $g_{\mu \nu}$, we obtain  the non--vacuum field equations
of $\mathrm{f(R)}$ gravitational theory as follows,
\cite{2005JCAP...02..010C}:
\begin{eqnarray} \label{f1}
{\mathop{\mathcal{ I}}}_{\mu \nu}=\mathit{ R}_{\mu \nu}
\mathrm{f_{R}}-\frac{1}{2}g_{\mu \nu}\mathrm{f( R)}+[g_{\mu
\nu}\Box -\nabla_\mu \nabla_\nu]\mathrm{ f}_{_\mathrm{ R}} -\kappa
T_{\mu \nu}\equiv0,\end{eqnarray} where  $\Box$ is the
d'Alembertian operator, $\displaystyle
\mathrm{f_{R}}=\frac{\mathrm {df}}{\mathrm {dR}}$ and the matter
energy--momentum tensor  $T_{\mu \nu}$ is defined as,
\begin{eqnarray} \label{Tmu}
T_{\mu \nu}=-\frac{2}{\sqrt{-g}}\frac{\delta {\cal L_M}}{\delta g^{\mu \nu}}.
  \end{eqnarray}
The trace of Eq.  ~(\ref{f1}), takes the following form,
\begin{eqnarray} \label{f3}
{\mathop{\mathcal{ I}}}=3\Box {\mathrm f_{R}}+\mathrm{
R}{f_{R}}-2\mathrm f(R) -\kappa T\equiv0 \,,\qquad \qquad
\textrm{where} \qquad \qquad T=T_\mu^\mu.\end{eqnarray} From Eq.
(\ref{f3}),  $\mathrm f(R)$  can be isolated to obtain the
following form,
\begin{eqnarray} \label{f3s}
\mathrm f(R)=\frac{1}{2}\big[3\Box {\mathrm f_{R}}+\mathrm{
R}{f_{R}}-\kappa T\Big]\,.\end{eqnarray} Using Eq. (\ref{f3s}) in
Eq. (\ref{f1}) we obtain the following \cite{Kalita:2019xjq},
\begin{eqnarray} \label{f3ss}
{\mathop{\mathcal{ I}}}_{\mu \nu}=\mathrm{ R}_{\mu \nu}
\mathrm{f_{R}}-\frac{1}{4}g_{\mu \nu}\mathrm{ R}\mathit{
f}_{_\mathrm{ R}}+\frac{1}{4}g_{\mu \nu}\Box\mathrm{ f}_{_\mathrm{
R}} -\nabla_\mu \nabla_\nu\mathrm{ f}_{_\mathrm{ R}}-\kappa(T_{\mu
\nu}-\frac{1}{4}g_{\mu \nu} T)  \,.\end{eqnarray} In this study,
we shall assume that the energy-momentum tensor, $T_{\mu \nu}$,
has the following specific form in order to achieve anisotropic
form,
\begin{eqnarray}
&&{ T}_\mu{}^\nu{}=(p_\bot+\rho)u_\mu u^\nu +p_\bot\delta_\mu{}^\nu+(p_r-p_\bot)\zeta_\mu \zeta^\nu,
\end{eqnarray}
where $u_\mu$ is the timelike vector defined as $u^\mu=[1,0,0,0]$,
and $\zeta_\mu$ is the unit spacelike vector in the radial
direction defined as $\zeta^\mu=[0,1,0,0]$ such that $u^\mu
u_\mu=-1$ and $\zeta^\mu\xi_\mu=1$. In this study, $\rho$
represents the energy-density, and $p_r$ and $p_\bot$ are the
radial and tangential pressures, respectively.

In the following sections,  we  apply the field equations, namely,
Eqs. (\ref{f3}) and (\ref{f3ss}) to a spherically symmetric
spacetime having two unknown functions.

\section{\bf  Stellar equations in the f(R) gravitational theory}\label{S3}

To study the non--vacuum field  Eqs. (\ref{f3}) and (\ref{f3ss})
we use  the following form of a spherically symmetric spacetime
having two unknown functions,
\begin{eqnarray} \label{met12}
& &  ds^2=-e^{\alpha(r)}dt^2+\frac{dr^2}{e^{\beta(r)}}+r^2d\Sigma\,, \qquad {\textrm where} \qquad d\Sigma=(d\theta^2+\sin^2d\phi^2)\,,  \end{eqnarray}
 where $\alpha(r)$ and $\beta(r)$ are unknown functions.   The Ricci scalar for the metric (\ref{met12}) takes the following form:
\begin{eqnarray} \label{Ricci}
  {\textit R(r)}=\frac{e^{-\beta}r[r\alpha'\beta'-2r\alpha''-r\alpha'^2-4\alpha'+4\beta'-4]+4}{2r^2}\,,
  \end{eqnarray}
where $\alpha\equiv \alpha(r)$, $\beta\equiv \beta(r)$,
$\alpha'=\frac{d\alpha}{dr}$, $\alpha''=\frac{d^2\alpha}{dr^2}$
and $\beta'=\frac{d\beta}{dr}$.  For the line--element
(\ref{met12}) the non-vanishing components of the field  Eqs.
 (\ref{f3}) and  (\ref{f3ss})  have the following forms:
 \begin{eqnarray} \label{fes}
&& {\mathfrak I}_t{}^t=\frac{e^{-\alpha}[e^{\alpha-\beta}\{Fr^2( 2\alpha''-\alpha'\beta'+\alpha'^2)+4rF[r\alpha'+r\beta'-1]+r^2[3\alpha'F'-2F''+F'\beta']-4rF'\}+4e^{\alpha}F]}{r^2}-8\pi \rho\,,\nonumber\\
&& {\mathfrak I}_r{}^r=\frac{e^{-\beta}[Fr^2( 2\alpha''-\alpha'\beta'+\alpha'^2)-4F[r\alpha'+r\beta'+1]-r^2[\alpha'F'-6F''+3F'\beta']-4rF'+4e^{\alpha}F]}{r^2}+8\pi P_r\,,\nonumber\\
&& {\mathfrak I}_\theta{}^\theta={\mathfrak I}_\phi{}^\phi=\frac{F[4-e^{-\beta}(4-2r^2\alpha''+r^2\alpha'\beta'-r^2\alpha'^2)]+e^{-\beta}[r^2\alpha'\beta'+2r^2F''-r^2F'\beta'-4rF']}{r^2}-8\pi P_\bot\,,\nonumber\\
&&{\mathfrak I}= [P_r+2P_\bot-\rho]+\frac{e^{-\beta}[r^2(6F''-2F\alpha''-F\alpha'^2)+r[rF\beta'+3rF'-4F]\alpha'+r\beta'[4F-3rF']+12rF'-4F)+4(F-f)}{8\pi r^2}\,,\nonumber\\
&&
\end{eqnarray}
where $F=\mathrm{ f}_{_\mathrm{ R}}=\frac{\mathrm{ df}}{\mathrm{
dR}}=\frac{\mathrm{ df}}{\mathrm{ dr}}\frac{\mathrm{ dr}}{\mathrm{
dR}}$.  The system of equations in (\ref{fes}) includes four
nonlinear differential equations with six unknown functions,
$\alpha$, $\beta$, $F$ $\rho$, $P_r$ and $P_t$; therefore, we must
impose two constraints to transform the equations in (\ref{fes})
into a closed system. In this study, we  use the Krori-Barua
ansatz that has the following  form \cite{Mustafa:2020yux}:
\begin{eqnarray} \label{mets}
  \alpha=b_0r^2+b_1\qquad \qquad \qquad \beta=b_2r^2\,,
  \end{eqnarray}
where $b_0$, and $b_2$ are the dimensionful parameters with the
inverse unit  of $r^2$, and $b_1$ is a constant.   Using Eq.
(\ref{mets}) in Eq. (\ref{fes}), we obtain the following:
 \begin{eqnarray} \label{sol}
 && \rho=\frac{e^{-b_0r^2}[b_0c_1(b_0-b_2)r^6+[a_0{}^2+(6c_1-b_2)b_0+3c_1b_2]r^4+(2b_2-4c_1+3b_0)r^2-1]+1+c_1r^2}{16\pi r^2}\,, \qquad  F=1+c_1r^2\,,\nonumber\\
 &&P_r(r)=\frac{e^{-b_2r^2}[1+(b_0+2b_2)r^2-r^4(b_0{}^2-b_0[2c_1+b_2]-5b_2c_1)+b_0c_1(b_2-b_0)r^6-e^{b_2r^2}(1+c_1r^2)]}{\pi r^2}\,,\nonumber\\
 && P_\bot(r)=\frac{1+c_1r^2-e^{-b_2r^2}[1+r^2(2c_1-b_0)-r^4[b_0(b_0+2c_1-b_2)-b_2c_1]+b_0c_1r^6(b_2-b_0)]}{16\pi r^2}\,.\nonumber\\
  \end{eqnarray}

\section{\bf  Matching conditions }\label{S4}

Given that solution ({\ref{sol}) has a nontrivial Ricci scalar as
shown in Eq. (\ref{Ris}), we must match it with an exterior
solution that has a non-constant Ricci scalar. In order to
exemplify our study and confront it with a realistic physical
system, we shall use the pulsar \textrm{Her X--1}, which has well
known mass and radius, whose estimated mass and radius are $M
=0.85\pm 0.15 M_\circledcirc$ and $b \thickapprox 8.1\pm0.41$ km,
respectively \cite{Gangopadhyay:2013gha}.

Thus, we  match solution (\ref{mets}), considering  $b_2=b_0$,
with the uncharged one presented in \cite{Nashed:2019tuk}. The
spherically symmetric uncharged solution  \cite{Nashed:2019tuk}
takes the following form
    \begin{eqnarray}\label{Eq1} ds^2= -\Big(\frac{1}{2}-\frac{2M}{r}\Big)dt^2+\Big(\frac{1}{2}-\frac{2M}{r}\Big)^{-1}dr^2+r^2d\Omega^2,
 \end{eqnarray}
where $M$ is the total mass of the stellar compact object and
$4M<r$. We have to match the interior spacetime metric
(\ref{mets}) with the exterior spacetime  given by Eq. (\ref{Eq1})
at the boundary of the star $r =b$. The continuity of the metric
functions across the boundary $r =b$ yields the following
conditions,
\begin{eqnarray}\label{Eq2} \alpha(r=b)=\Bigg(\frac{1}{2}-\frac{2M}{b}\Bigg), \qquad \qquad \beta(r=b)=\Bigg(\frac{1}{2}-\frac{2M}{b}\Bigg)^{-1}.
 \end{eqnarray}
Using the above conditions we get the constraints on the constants
$b_0$, $b_1$. The functional form of these constants takes the
form,
\begin{eqnarray}\label{Eq3} b_0=\frac{ln\Big(\frac{2b}{b-4M}\Big)}{b^2}, \qquad \qquad b_1=ln\Big(\frac{b^2-8bM+16M^2}{4b^2}\Big)\,.\end{eqnarray} In  Figs.  \ref{Fig:1}   \ref{Fig:2} we plot  the metric potentials and  matching metric, respectively.
\begin{figure}
\centering
\subfigure[~Metric $g_{tt}$]{\label{fig:pot1}\includegraphics[scale=0.3]{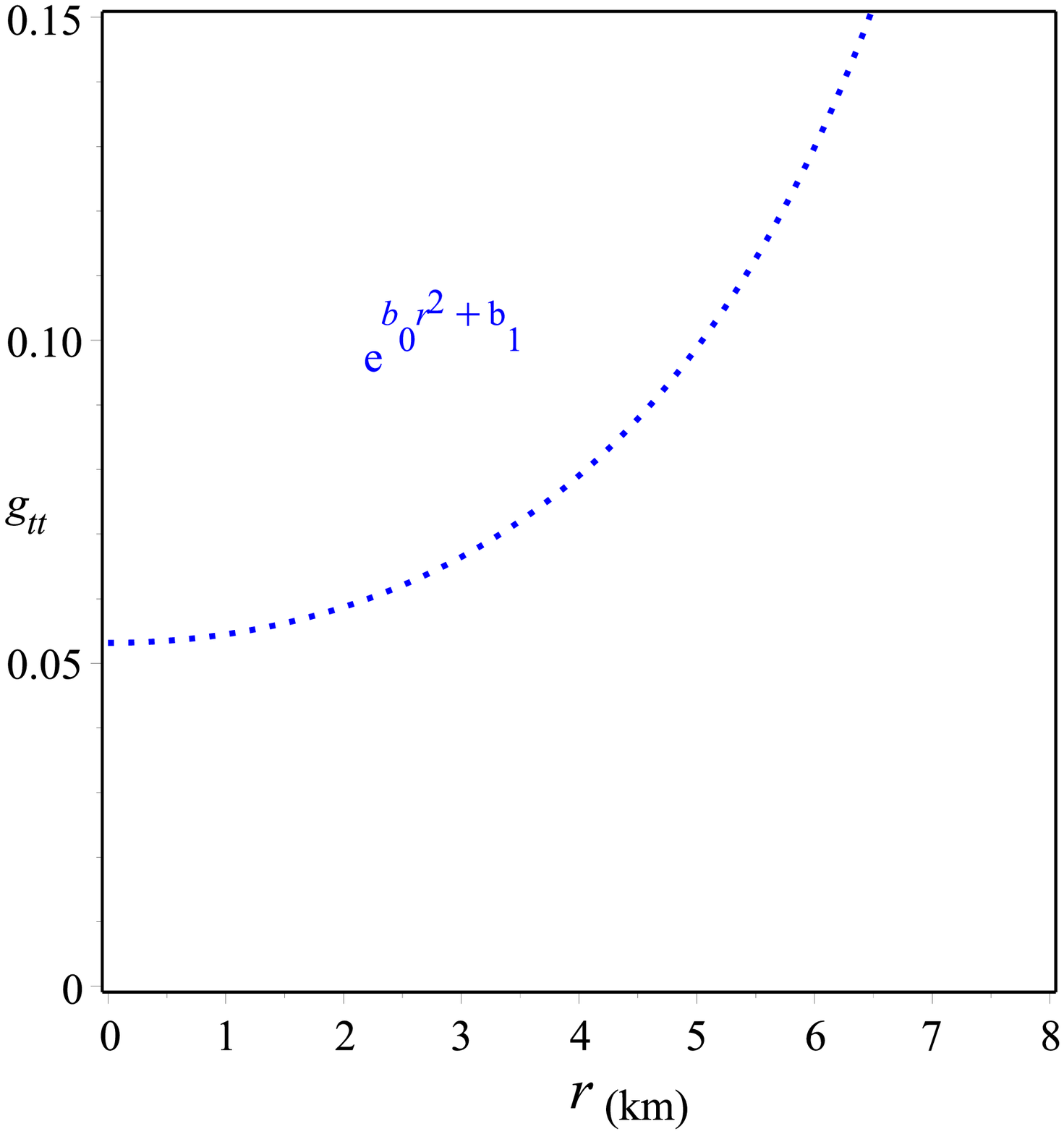}}
\subfigure[~Metric $g_{rr}$]{\label{fig:pot2}\includegraphics[scale=.3]{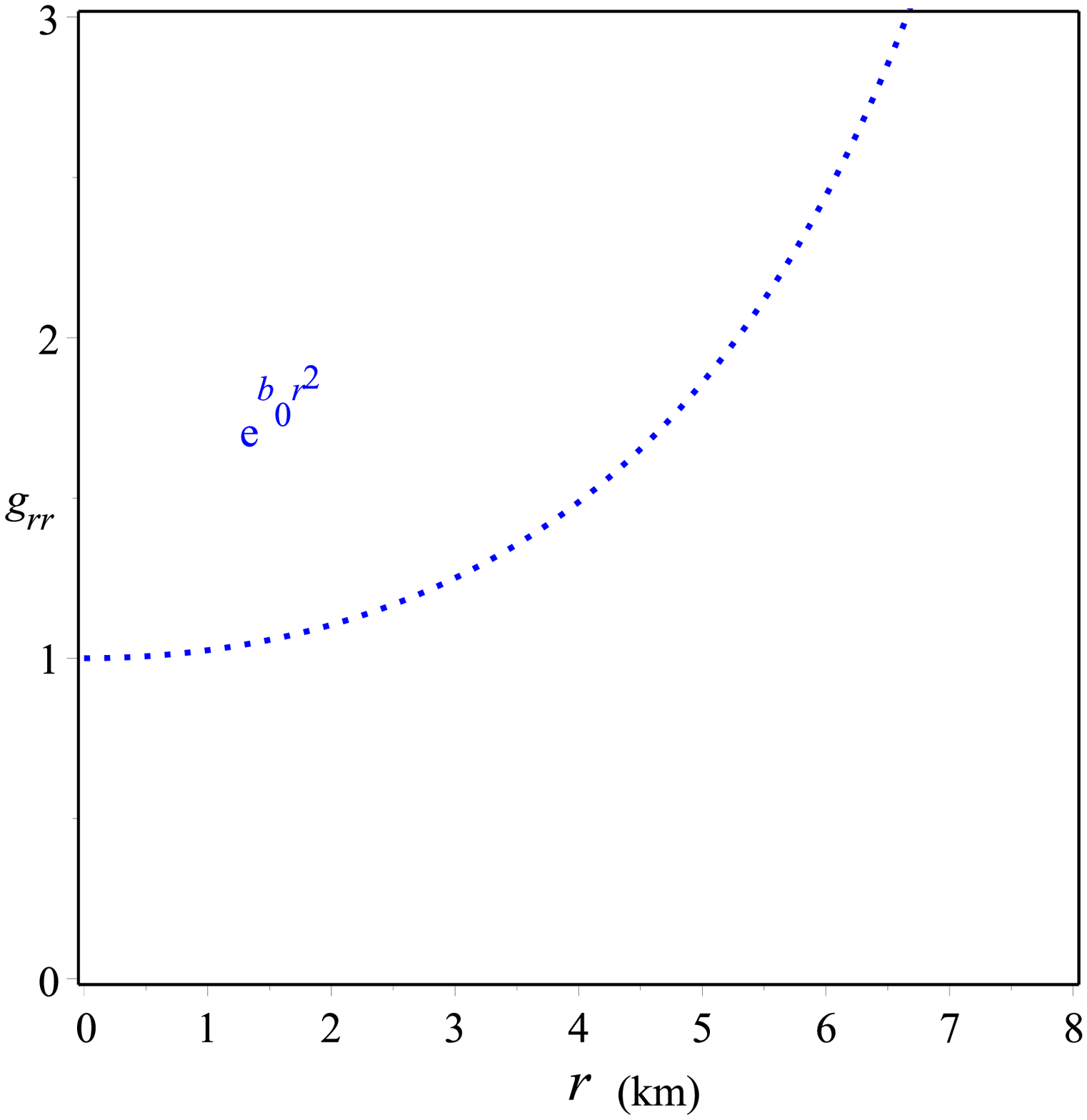}}
\caption[figtopcap]{\small{{Schematic plot of the radial coordinate $r$ in Km versus the potentials of the metric (\ref{mets}) using the constants constrained from RX J 1856--37 from where we put $b_0=0.0259974$ and $b_1=-3.7654625$. These values of the two constants $b_0$ and $b_1$ will be use throughout this study.}}}
\label{Fig:1}
\end{figure}
\begin{figure}
\centering
\subfigure[~Matching condition of the potential $e^{\alpha(b)}$;]{\label{fig:sm}\includegraphics[scale=0.3]{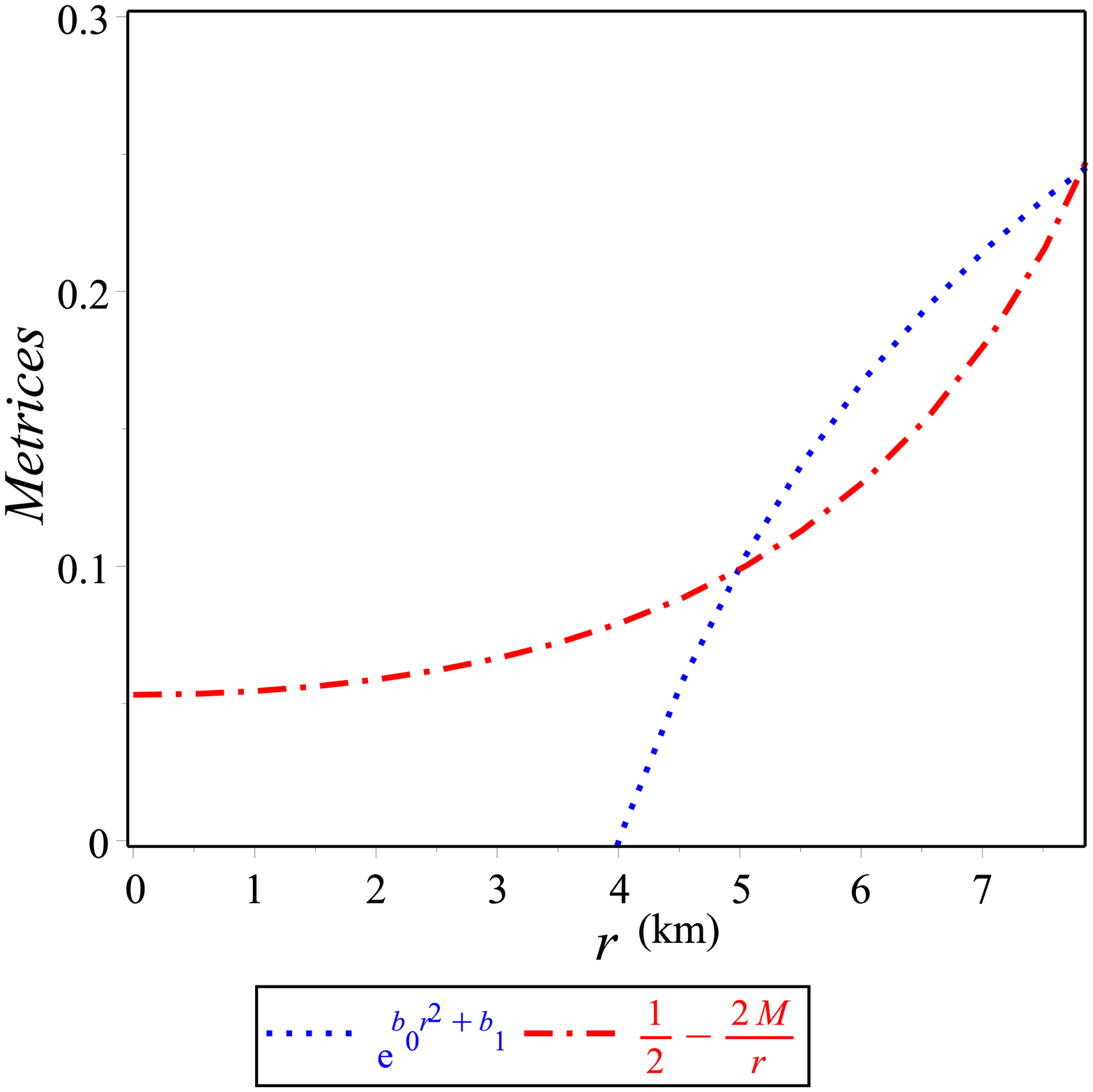}}
\subfigure[~matching condition of the potential $e^{\beta(b)}$]{\label{fig:sm1}\includegraphics[scale=.3]{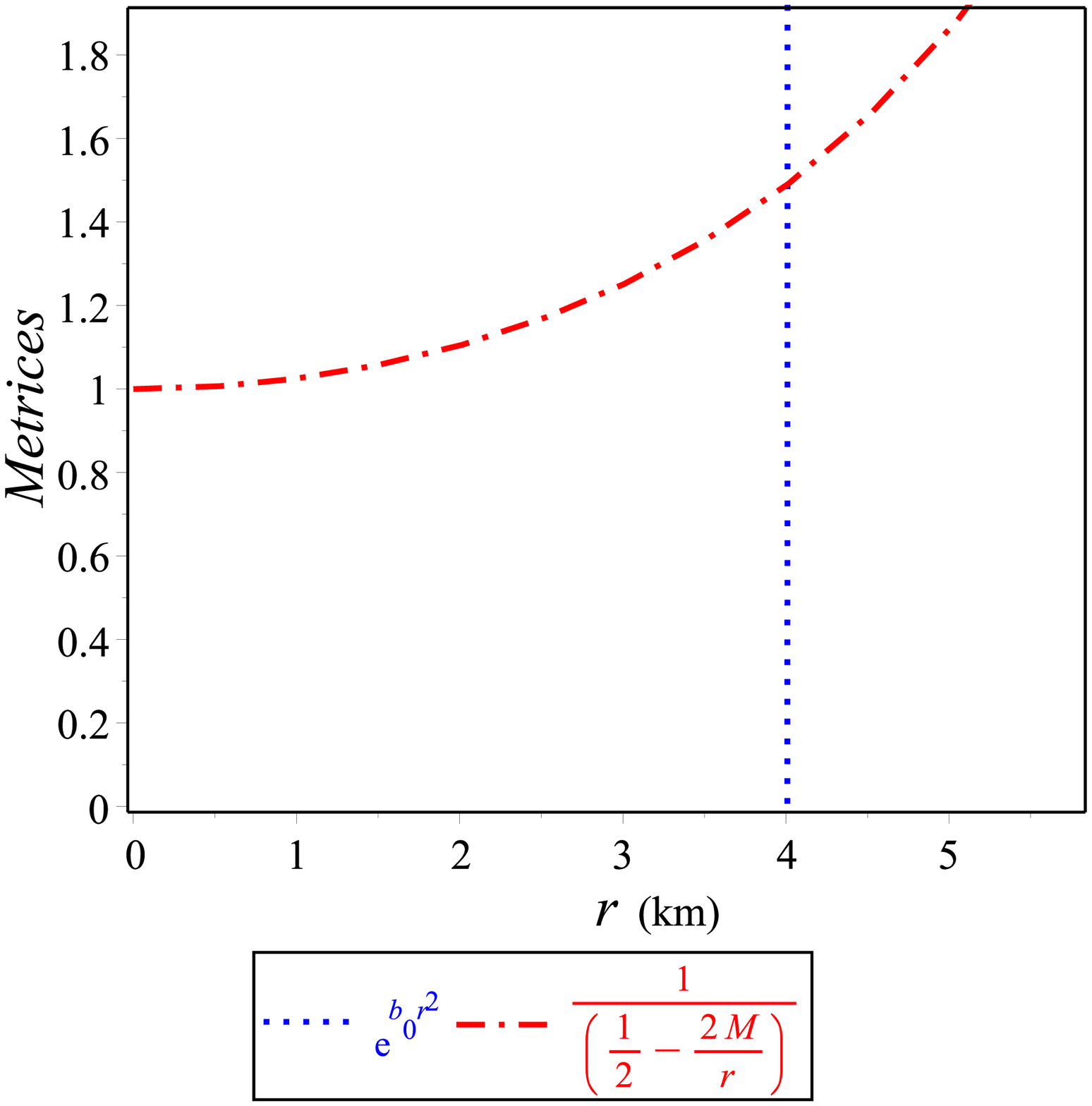}}
\caption[figtopcap]{\small{{Plot of boundary matching  of  \subref{fig:sm} $g_{tt}$ and  \subref{fig:sm1} $g_{rr}$ of (\ref{sol}).}}}
\label{Fig:2}
\end{figure}

\section{\bf   Terms of physical  viability of the solution (\ref{sol}) }\label{S5}

To investigate whether  the interior solution (\ref{sol}) is
suitable to describe a physical system, several criteria must be
satisfied, thus, we explore these criteria in this section.

\subsection{\bf Energy--momentum tensor}

For a realistic  interior solution,  positive values of
energy--density, radial, and transverse  pressures are needed.
Furthermore, all these quantities have finite values at the center
of the star. These energies gradually  decrease  toward the
surface of the star, and $P_r\geq P_\bot$. Fig. \ref{Fig:3} shows
the portraits of the martial energy, density, radial, and
transverse pressures. The figure shows that,
$\rho(r=0)_{c_1=0}=0.003104779987$,
$\rho(r=0)_{c_1=-0.01}=0.003701913745$,
$\hat{p}_r(r=0)_{c_1=0}=0.001034926662$,
$\hat{p}_r(r=0)_{c_1=-0.01}=0.001233971248$,
$\hat{p}_\bot(r=0)_{c_1=0}=0.001034926662$,
$\hat{p}_\bot(r=0)_{c_1=0}=0.001233971248$.  Fig.  \ref{Fig:3} also
displays  that all the components of the energy-momentum tensor
gradually decrease toward the surface of the star. From Fig.
\ref{Fig:3}, values of  energy--density, radial and transverse
pressures at the center in the case of $c_1=0$ are smaller than
those when  $c_1\neq0$. Moreover the components of the
energy-momentum proceed to the surface of the star more rapidly at
$c_1=-0.01$ than at  $c_1=0$ and $P_r=P\bot$ at the center.
However, as we approach the surface of the star   $\hat{p}_\bot\geq
P_r$. This behavior  is illustrated in  Fig. \ref{Fig:4}, which
also shows anisotropy behavior that is defined as
$\Delta(r)=\hat{p}_\bot-\hat{p}_r$. Fig.  \ref{Fig:4} also reveals that the
anisotropic force is positive, which means that it is a repulsive
force because $P_\bot\geq P_r$.

Moreover, the gradient of the density, radial, and transverse
pressures must be negative inside the stellar body, i.e.,
$\frac{d\rho}{dr}< 0$, $\frac{d\hat{p}_r}{dr}< 0$ and
$\frac{d\hat{p}_\bot}{dr}< 0$ \cite{Chanda:2019hyh}. Using (\ref{sol}),
we calculate the derivative of density, radial, and transverse
pressures as,
\begin{eqnarray} \label{grad}
 && \rho'=\frac{d\rho}{dr}\nonumber\\
 &&=\frac{e^{-b_0r^2}[1-b_0b_2c_1(b_0-b_2)r^8-r^6([3c_1-b_0]b_2{}^2+[8b_0c_1+b_0{}^2]b_2-2c_1b_0{}^2)-r^4(2b_2{}^2+(4b_0-7c_1)b_2-b_0{}^2-6b_0c_1)]-1}{r^3}\,,\nonumber\\
 &&P'_r=\frac{dP_r}{dr}\nonumber\\
 &&
 =\frac{e^{-b_0r^2}[b_0b_2c_1(b_0-b_2)r^8-r^6([b_0+5c_1]b_2{}^2-[b_0b_2-2b_0c_1]b_0)-r^4(b_0{}^2-5b_2c_1+2b_2{}^2-2b_0c_1)-b_2r^2-1]+1}{r^3}\,,\nonumber\\
 && P'_\bot=\frac{dP_\bot}{dr}\nonumber\\
 &&
 =\frac{e^{-b_0r^2}[1-b_0b_2c_1(b_0-b_2)r^8+r^6([2c_1-b_2]b_0{}^2-[4c_1-b_2]b_2+b_2{}^2c_1)+r^4(b_0{}^2-2b_0[b_2+c_1]+b_2c_1)+b_2r^2]-1}{r^3}\,.\nonumber\\
  \end{eqnarray}
The behavior of the gradients of  density, radial, and transverse
pressures are shown in Fig. \ref{Fig:5}, where it can also be seen
that $\rho'$, $\hat{p}'_r$ and  $\hat{p}'_\bot$ have negative values as
required by a real stellar compact object.
\begin{figure}
\centering
\subfigure[~Density of solution  (\ref{sol})]{\label{fig:dnesity}\includegraphics[scale=0.3]{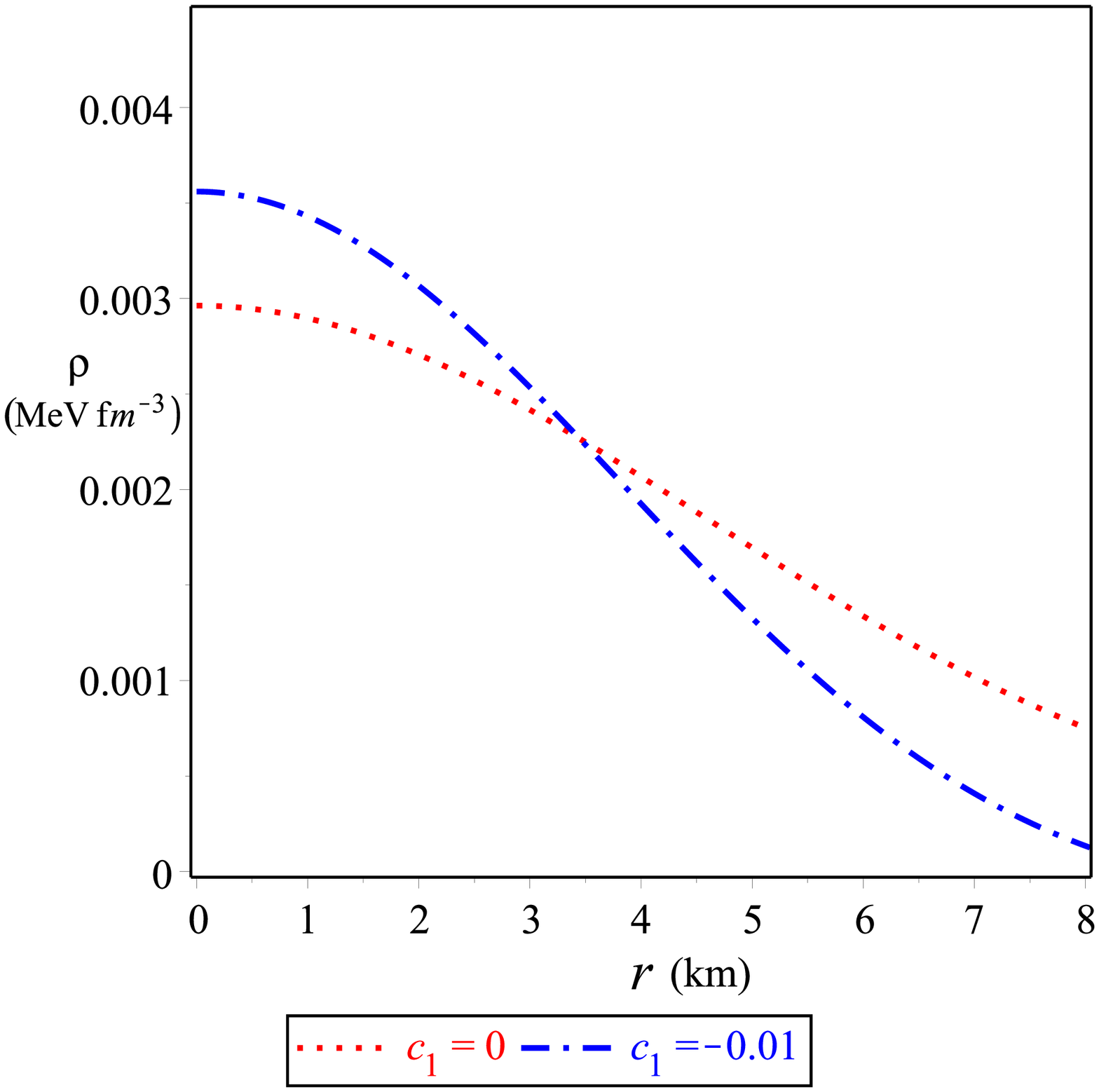}}
\subfigure[~Radial pressure of solution  (\ref{sol})]{\label{fig:pressure}\includegraphics[scale=.3]{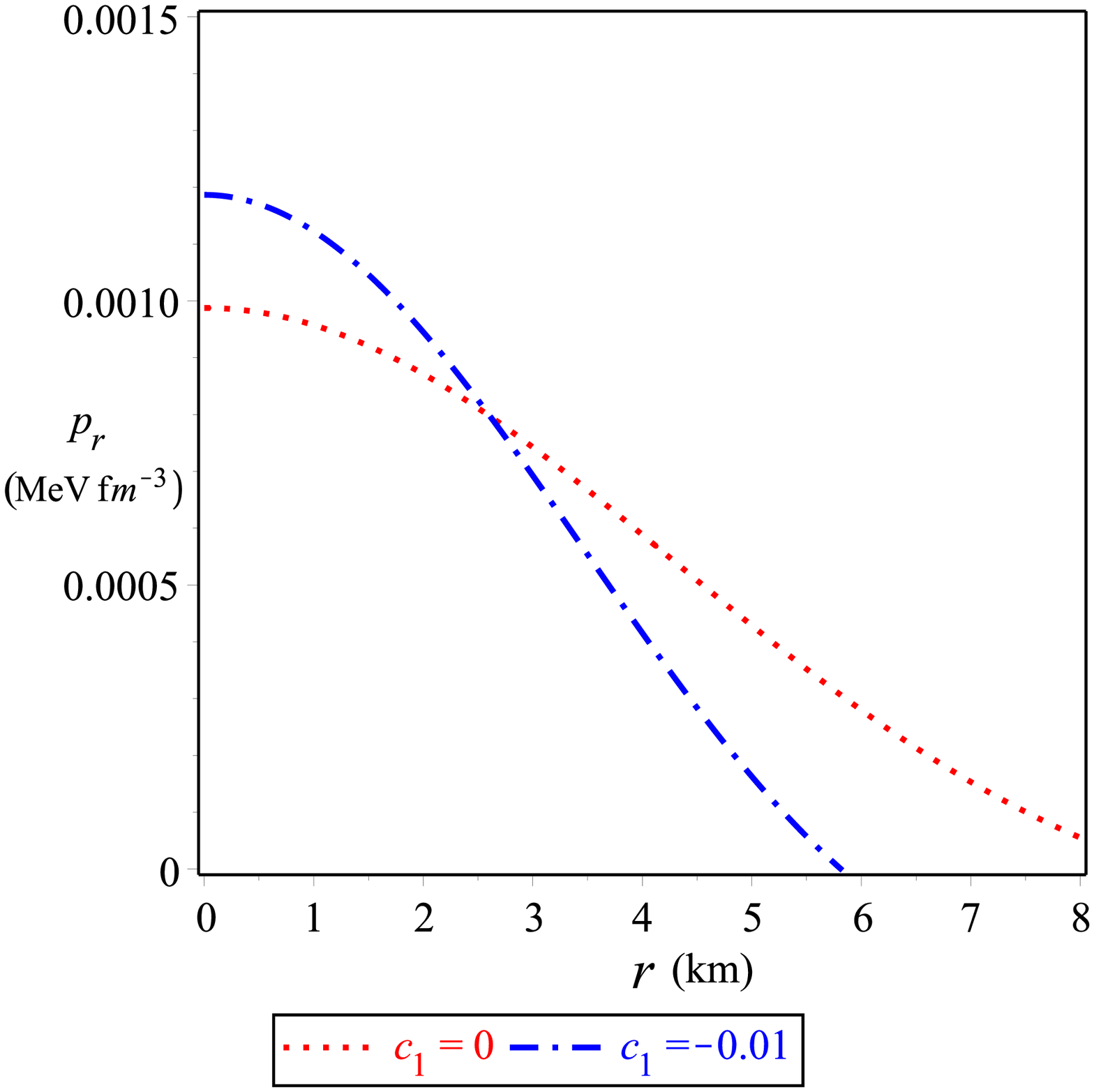}}
\subfigure[~Transverse pressure of solution  (\ref{sol})]{\label{fig:EoS}\includegraphics[scale=.3]{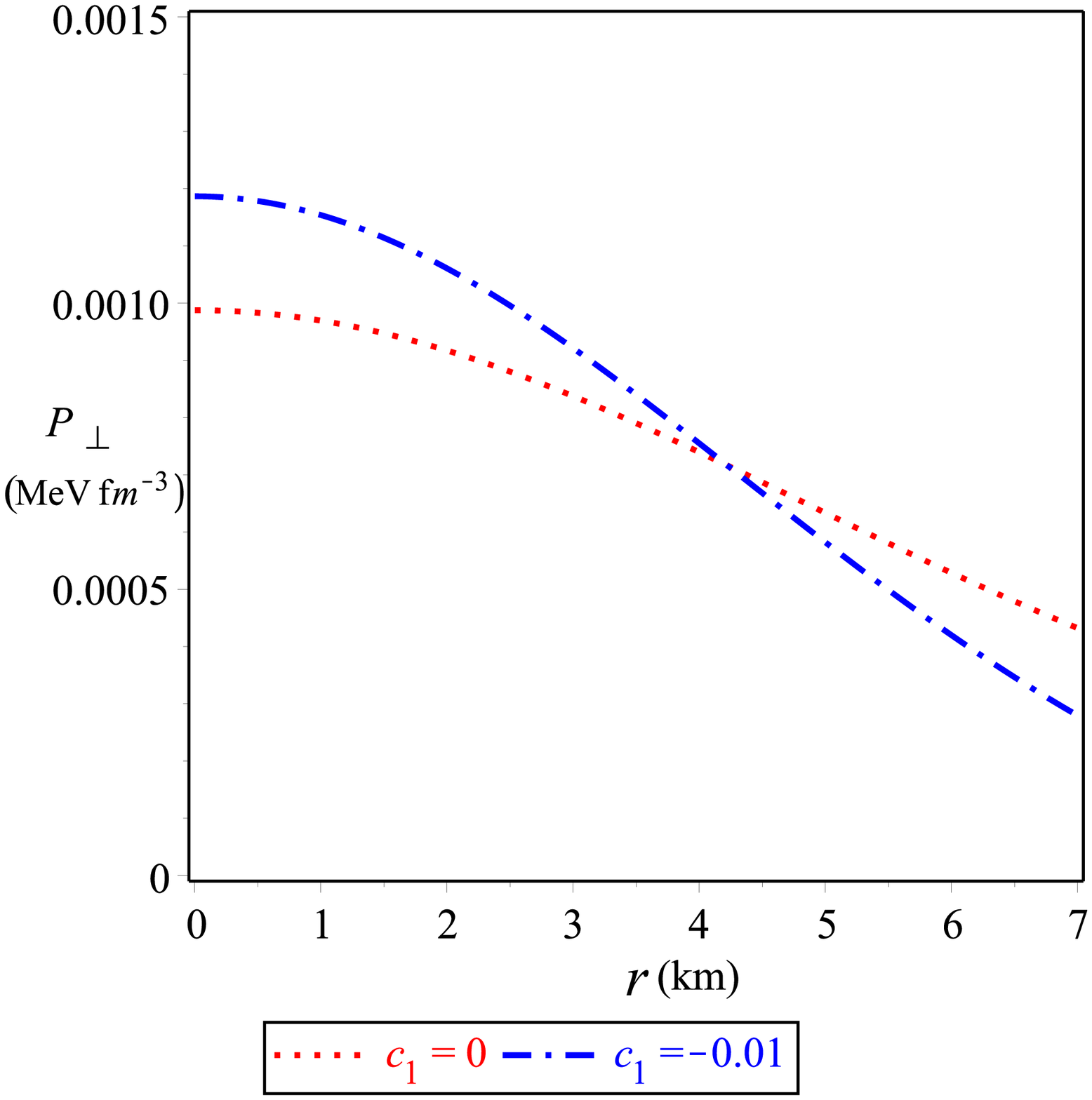}}
\caption[figtopcap]{\small{{Plots of density, radial, and transverse pressures. All the plots show  that the components of the energy-momentum tensors are positive as required by any real stellar.}}}
\label{Fig:3}
\end{figure}

\begin{figure}
\centering
\subfigure[~Anisotropy $\Delta$  of the solution (\ref{sol})]{\label{fig:Ani}\includegraphics[scale=0.3]{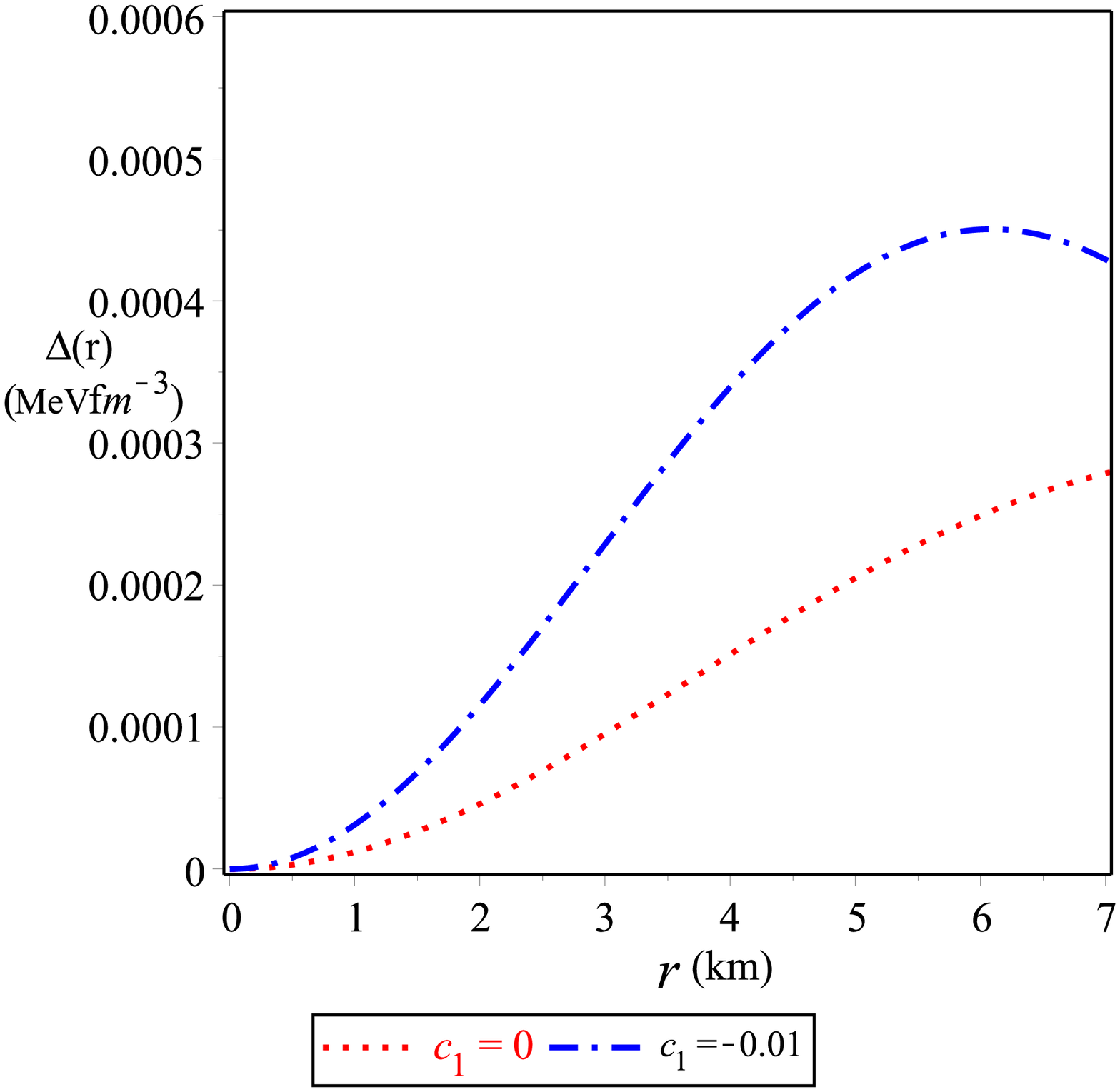}}
\subfigure[~Anisotropic force for $\Delta/r$  of solution (\ref{sol})]{\label{fig:pressure}\includegraphics[scale=.3]{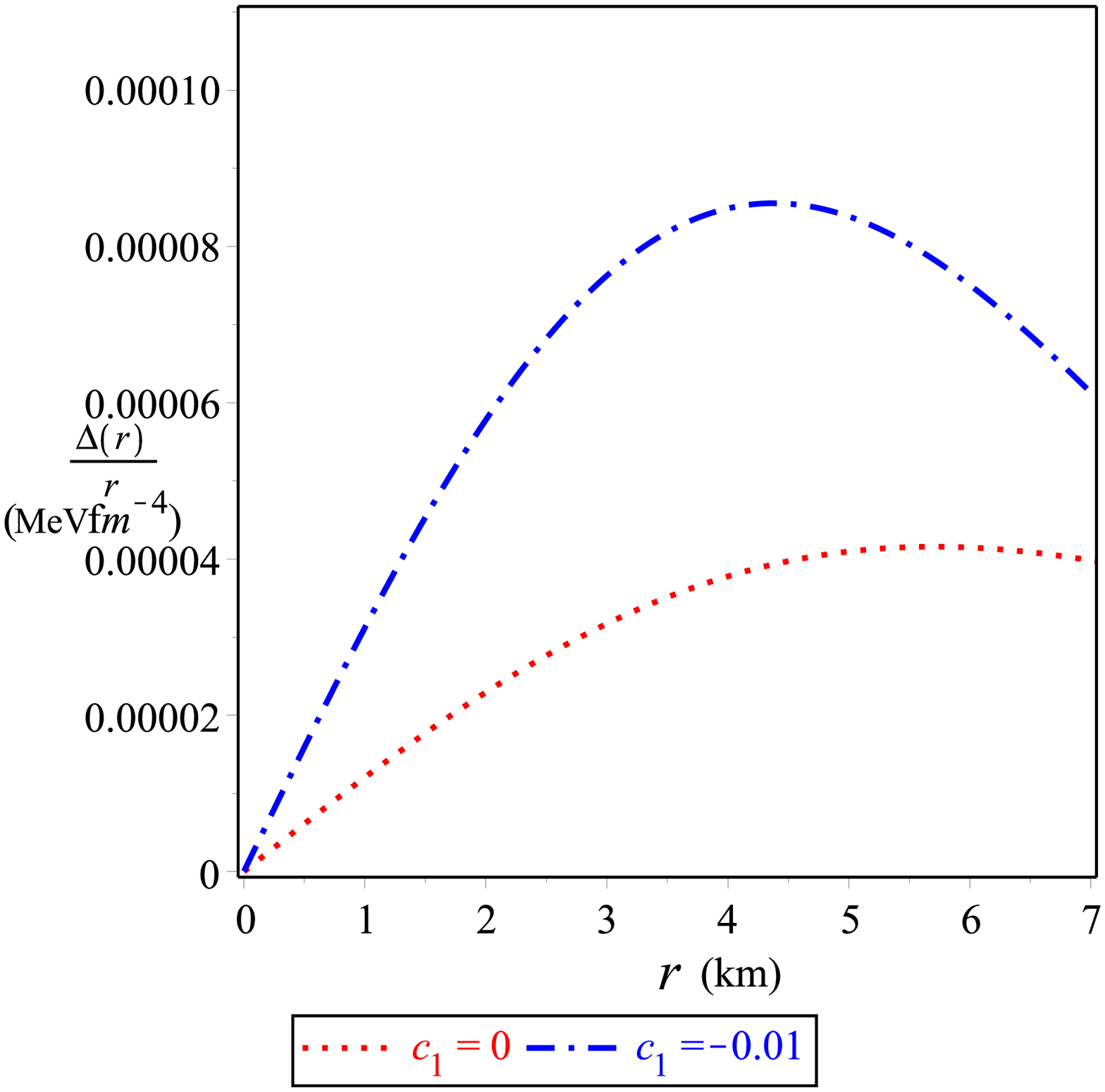}}
%
\caption[figtopcap]{\small{{Anisotropy $\Delta(r)$ and anisotropy force. Plot \subref{fig:Ani} shows that we have a repulsive force due to the positivity of the anisotropy. }}}
\label{Fig:4}
\end{figure}

\begin{figure}
\centering
\subfigure[~Gradient of density of solution (\ref{sol})]{\label{fig:dnesity}\includegraphics[scale=0.3]{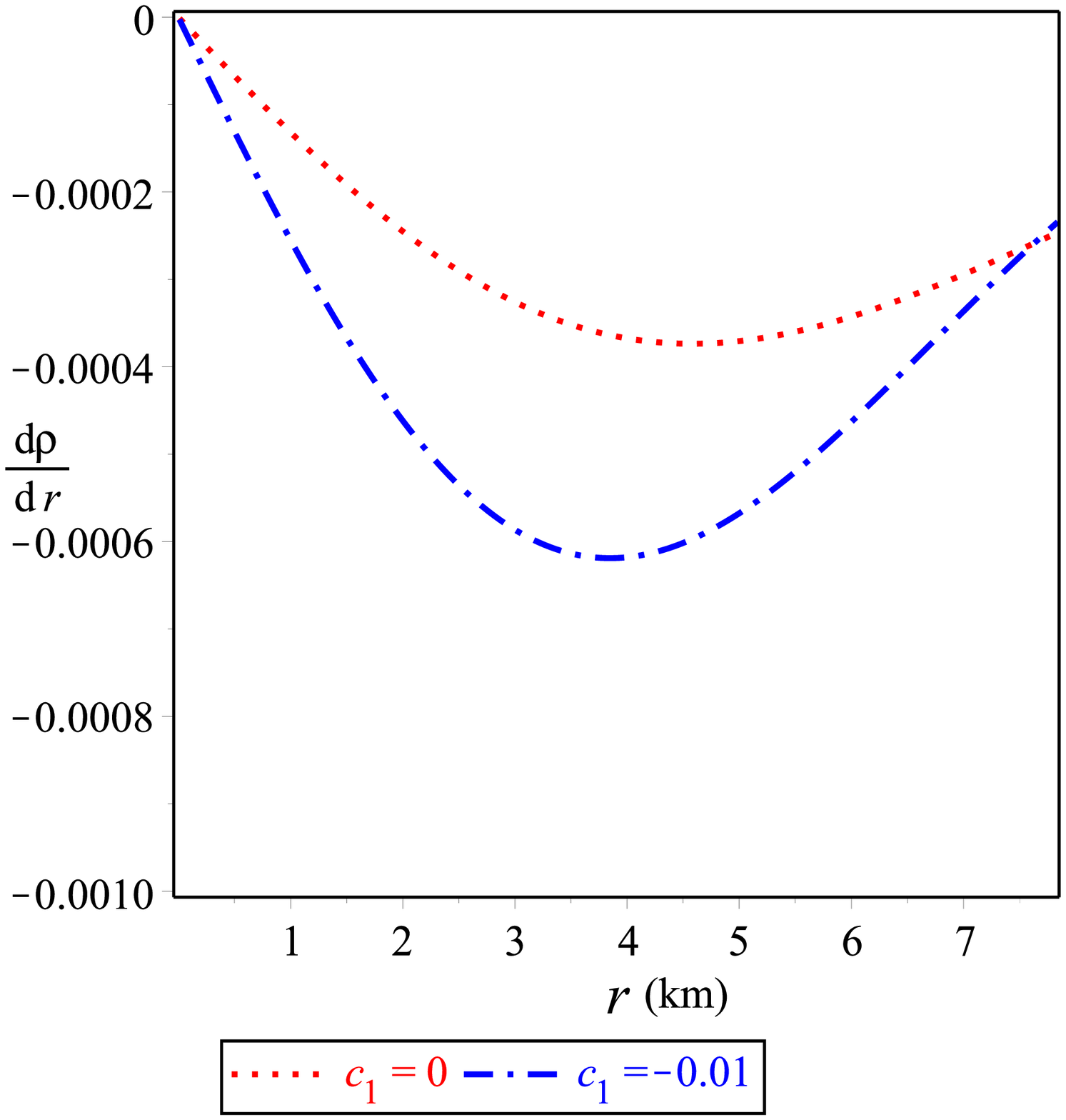}}
\subfigure[~Gradient of radial pressure  of solution (\ref{sol})]{\label{fig:pressure}\includegraphics[scale=.3]{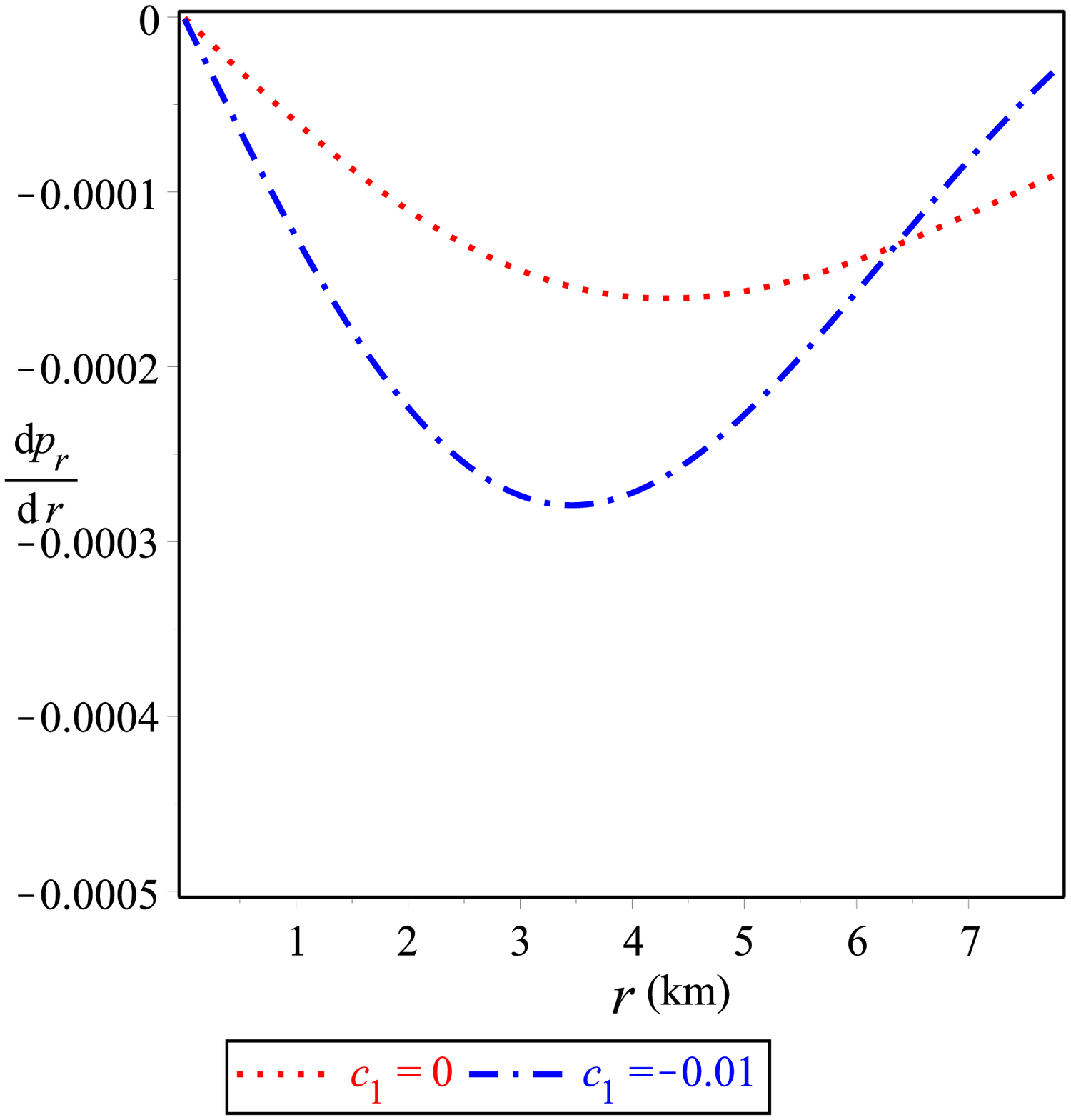}}
\subfigure[~Gradient of transverse pressure of solution  (\ref{sol})]{\label{fig:EoS}\includegraphics[scale=.3]{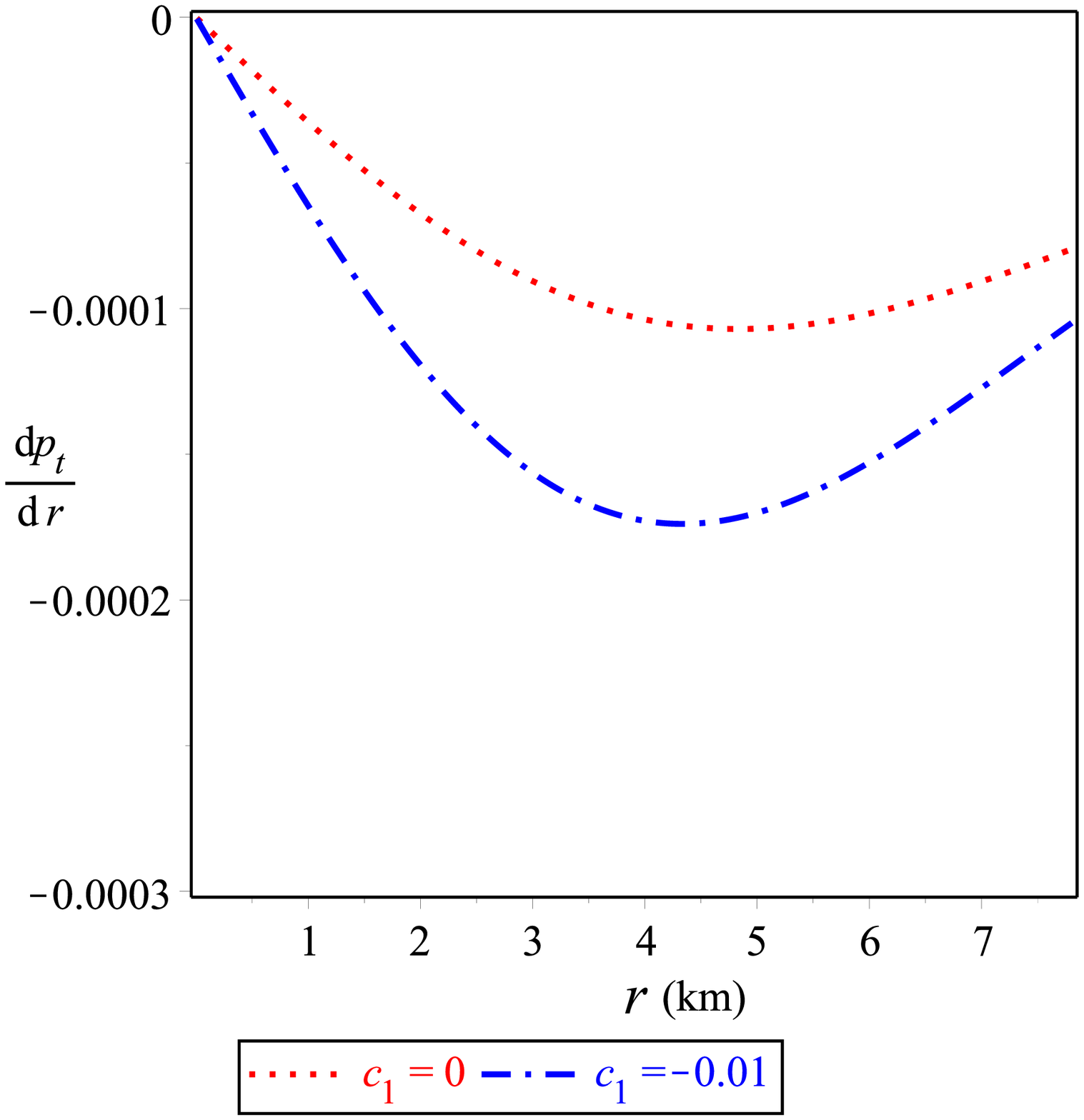}}
\caption[figtopcap]{\small{{Gradients of density, radial, and transverse pressures. All figures show  that the gradients of the components of the energy--momentum tensor are negative consistent with those of a  real stellar.}}}
\label{Fig:5}
\end{figure}

\subsection{\bf Causality}

To show the behavior of   sound velocities, we must calculate the
gradient of  energy--density, radial, and transverse  pressures
with the form given by Eq. (\ref{grad}).   Using Eq. (\ref{grad}),
we obtain the following:
  \begin{eqnarray} \label{grad}
 && v_r{}^2=\frac{P'_r}{\rho'}\,,\nonumber\\
 && =
 \frac{e^{-b_0r^2}[b_0b_2c_1(b_0-b_2)r^8-r^6([b_0+5c_1]b_2{}^2-[b_0b_2-2b_0c_1]b_0)-r^4(b_0{}^2-5b_2c_1+2b_2{}^2-2b_0c_1)-b_2r^2-1]+1}{e^{-b_0r^2}[1-b_0b_2c_1(b_0-b_2)r^8-r^6([3c_1-b_0]b_2{}^2+[8b_0c_1+b_0{}^2]b_2-2c_1b_0{}^2)-r^4(2b_2{}^2+(4b_0-7c_1)b_2-b_0{}^2-6b_0c_1)]-1}\,,\nonumber\\
 && v_\bot{}^2=\frac{P'_\bot}{\rho'}\,,\nonumber\\
 && =
 \frac{e^{-b_0r^2}[1-b_0b_2c_1(b_0-b_2)r^8+r^6([2c_1-b_2]b_0{}^2-[4c_1-b_2]b_2+b_2{}^2c_1)+r^4(b_0{}^2-2b_0[b_2+c_1]+b_2c_1)+b_2r^2]-1}{e^{-b_0r^2}[1-b_0b_2c_1(b_0-b_2)r^8-r^6([3c_1-b_0]b_2{}^2+[8b_0c_1+b_0{}^2]b_2-2c_1b_0{}^2)-r^4(2b_2{}^2+(4b_0-7c_1)b_2-b_0{}^2-6b_0c_1)]-1}\,.\nonumber\\
 \end{eqnarray}
To ensure that the causality condition is satisfied both for
radial and the transverse sound speeds, we must show that the
values of $v_r{}^2$ and $v_\bot{}^2$  are less than the speed of
light. To this end, we plot them in Fig.  \ref{Fig:6}  to ensure
that both of variables have values less than  the speed of light,
provided that the speed of light is unity in relativistic units.

{
Herrera assumed the cracking condition of a stable anisotropic compact star that results when equilibrium is
disturbed could be due to  local anisotropy. This condition is depend on the  radial and tangential
sound speeds, $v_r$ and $v_t$. Using Herrera condition \cite{1994PhLA..188..402H, Abreu:2007ew} that demonstrated that a simple
requirement in order to avoid gravitational cracking  is $-1\leq
v_t{}^2-v_r{}^2\leq0$. In Fig. \ref{Fig:6} \subref{fig:crac}, we
show that solution (\ref{sol}) is stable against cracking for
$c_1=0$ and $c_1=-0.01$. }
\begin{figure}
\centering
\subfigure[~Radial speed  of solution  (\ref{sol})]{\label{fig:vr}\includegraphics[scale=0.3]{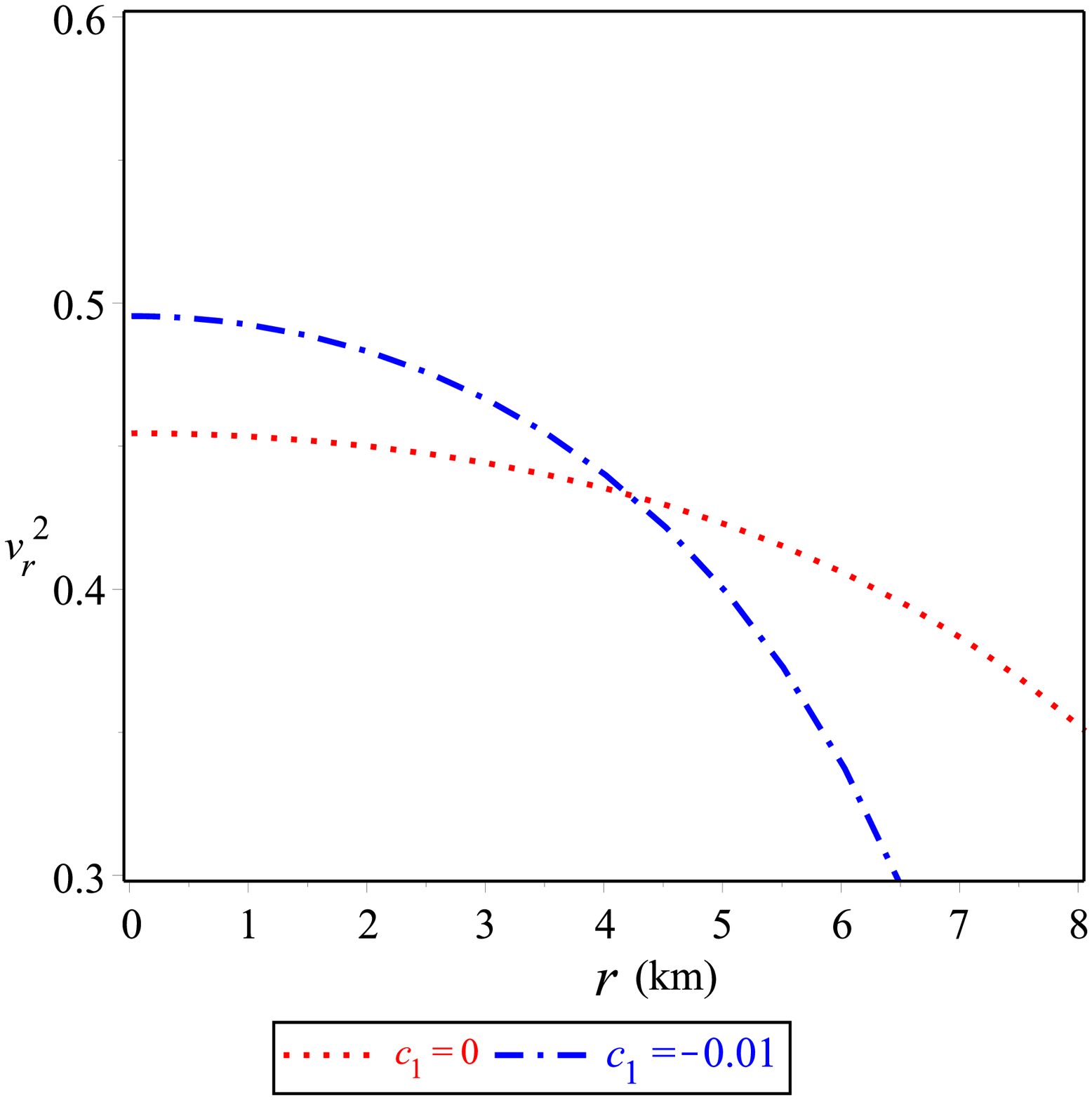}}
\subfigure[~Transverse speed  of solution  (\ref{sol})]{\label{fig:vt}\includegraphics[scale=.3]{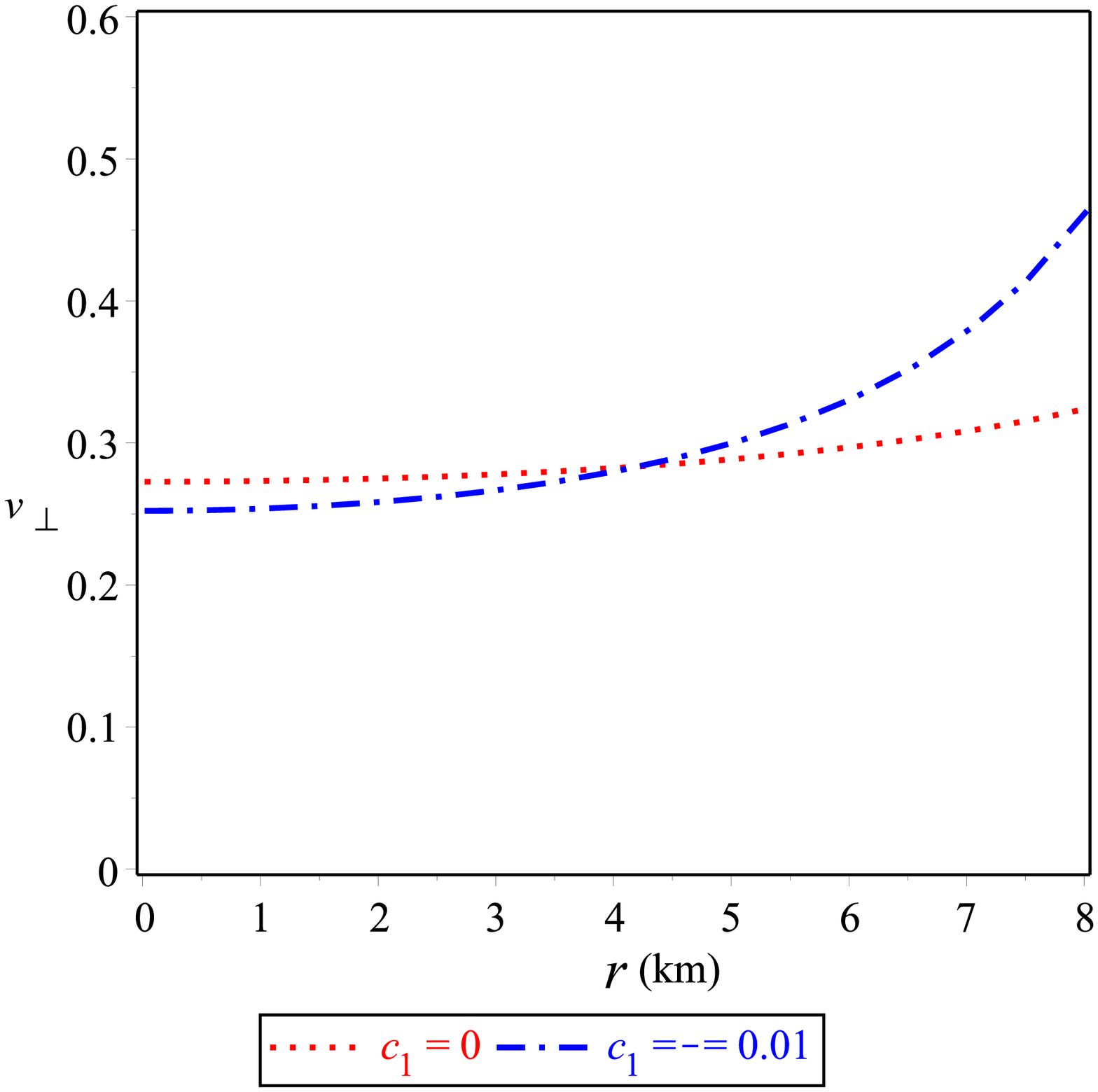}}
\subfigure[~Variation of $v_\bot{}^2-v_r{}^2$ of solution  (\ref{sol})]{\label{fig:crac}\includegraphics[scale=.3]{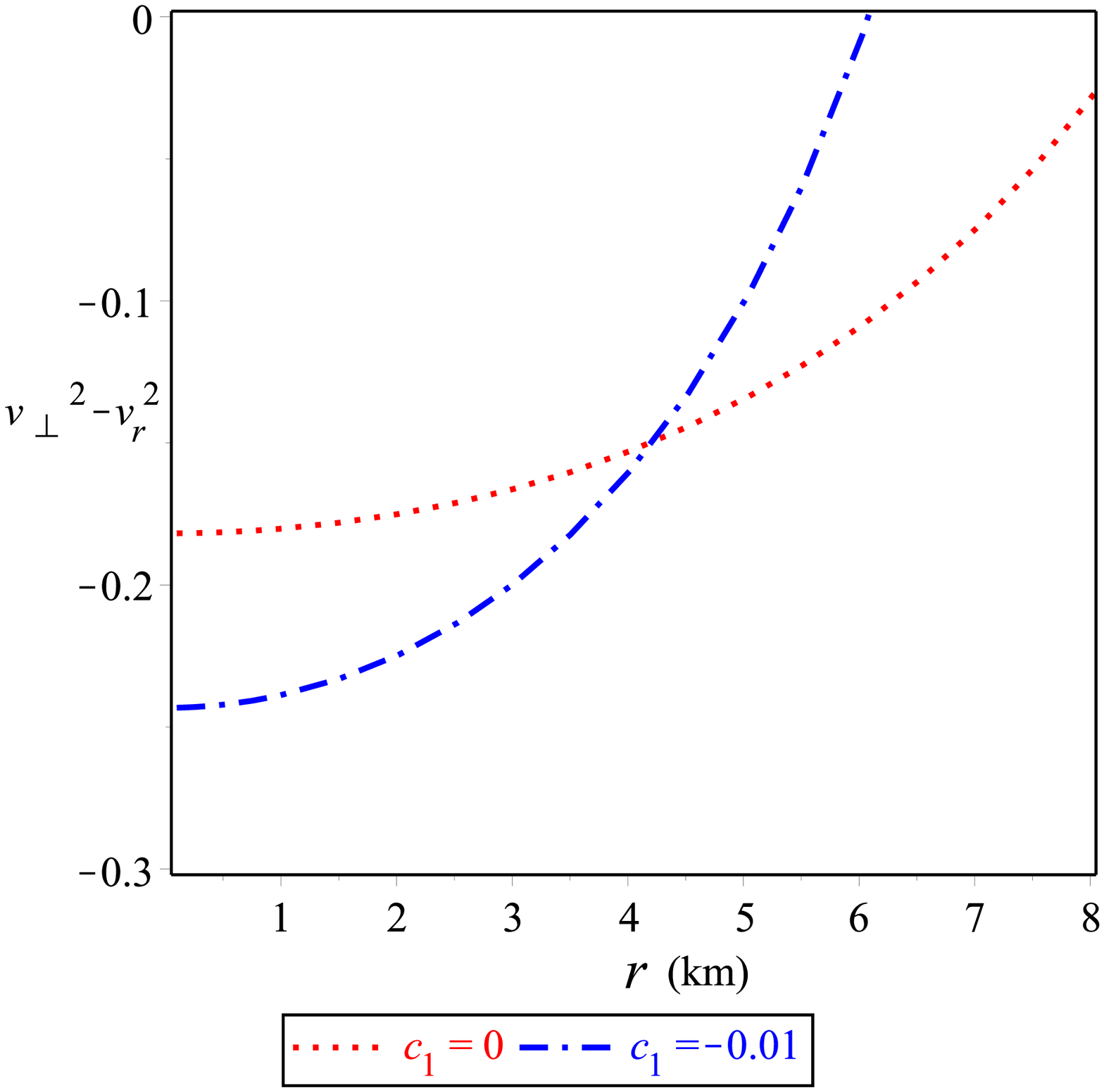}}
\caption[figtopcap]{\small{{Radial, transverse  speed of sound,  and    $v_\bot{}^2-v_r{}^2$. The plots of radial and transverse speeds indicate that our model   satisfies the causality condition. }}}
\label{Fig:6}
\end{figure}

\subsection{\bf Energy conditions}

For the non--vacuum solution, the energy conditions are considered
important tools. Therefore,  the dominant energy condition (DEC)
implies that the speed of energy should be less than the speed of
light.  To fulfill the DEC, we must have $\rho-P_r >0$ and
$\rho-P_\bot>0$. We show that the DEC is fulfilled  in  Fig.
\ref{Fig:7} moreover, we study the weak energy condition (WEC),
$\rho+P_r>0$ and $\rho+P_\bot>0$, and the strong energy condition
(SEC),  $\rho-P_r-2P_\bot>0$, and show in  Fig.  \ref{Fig:8}  that
both are satisfied.
\begin{figure}
\centering
\subfigure[~ DEC  $(\rho-P_r$),  (\ref{sol}),]{\label{fig:dnesity}\includegraphics[scale=0.3]{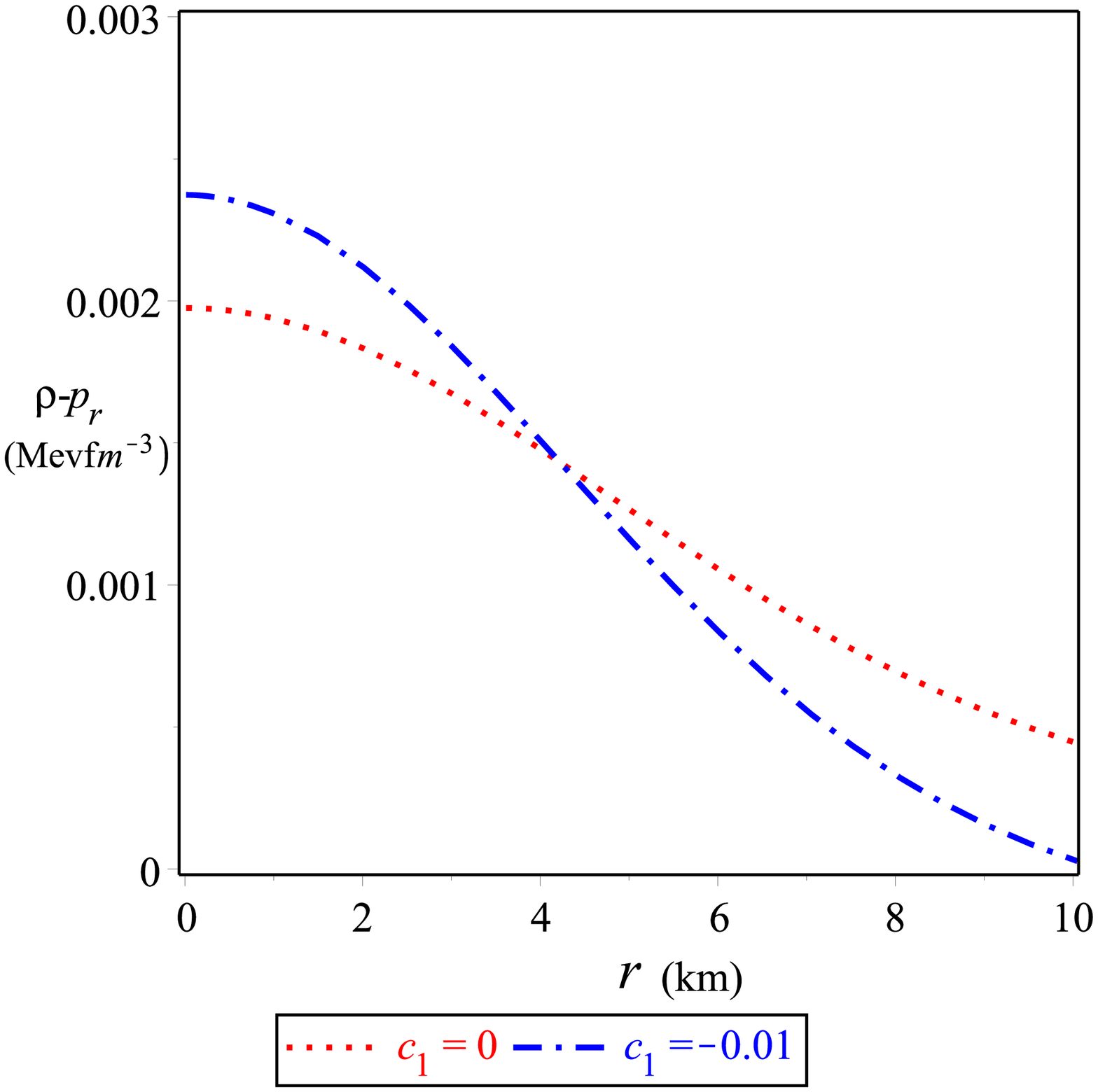}}
\subfigure[~ DEC $(\rho-P_\bot)$,  (\ref{sol})]{\label{fig:pressure}\includegraphics[scale=.3]{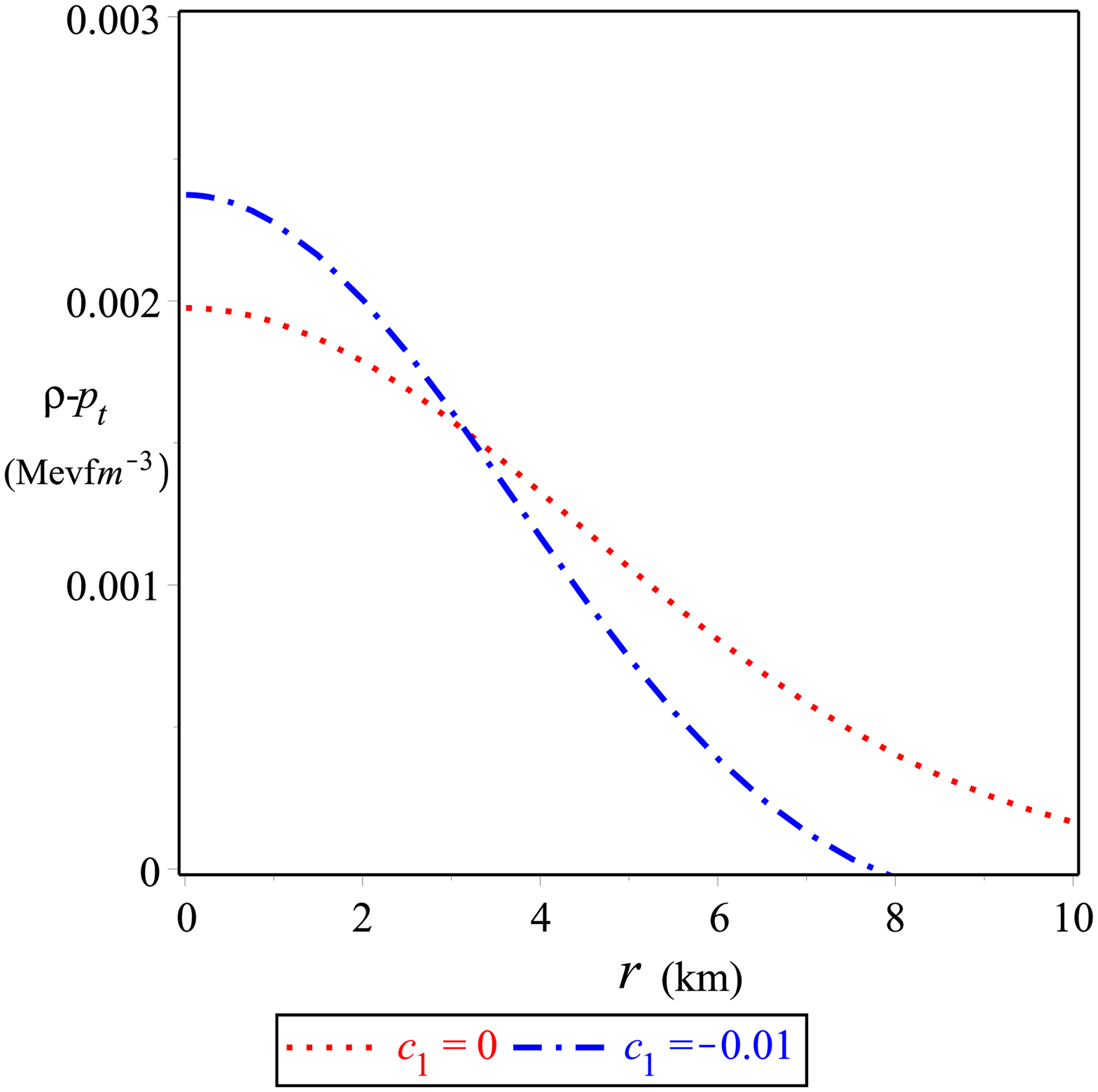}}
%
\caption[figtopcap]{\small{{DEC of solution  (\ref{sol}).}}}
\label{Fig:7}
\end{figure}
\begin{figure}
\centering
\subfigure[~ WEC  $(\rho+P_r)$   (\ref{sol}),]{\label{fig:dnesity}\includegraphics[scale=0.3]{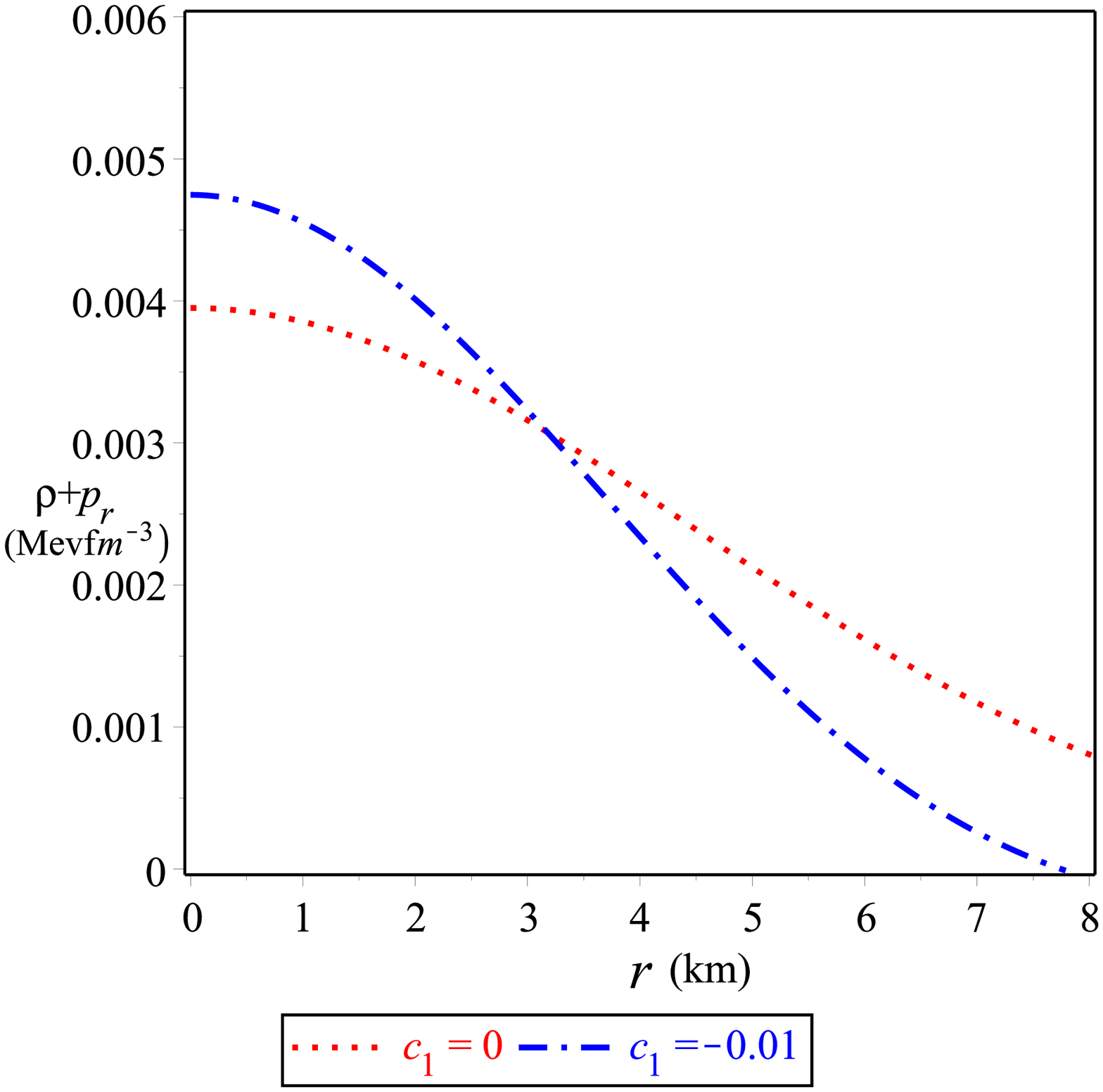}}
\subfigure[~ WEC  $(\rho+P_\bot)$  (\ref{sol}) and ]{\label{fig:pressure}\includegraphics[scale=.3]{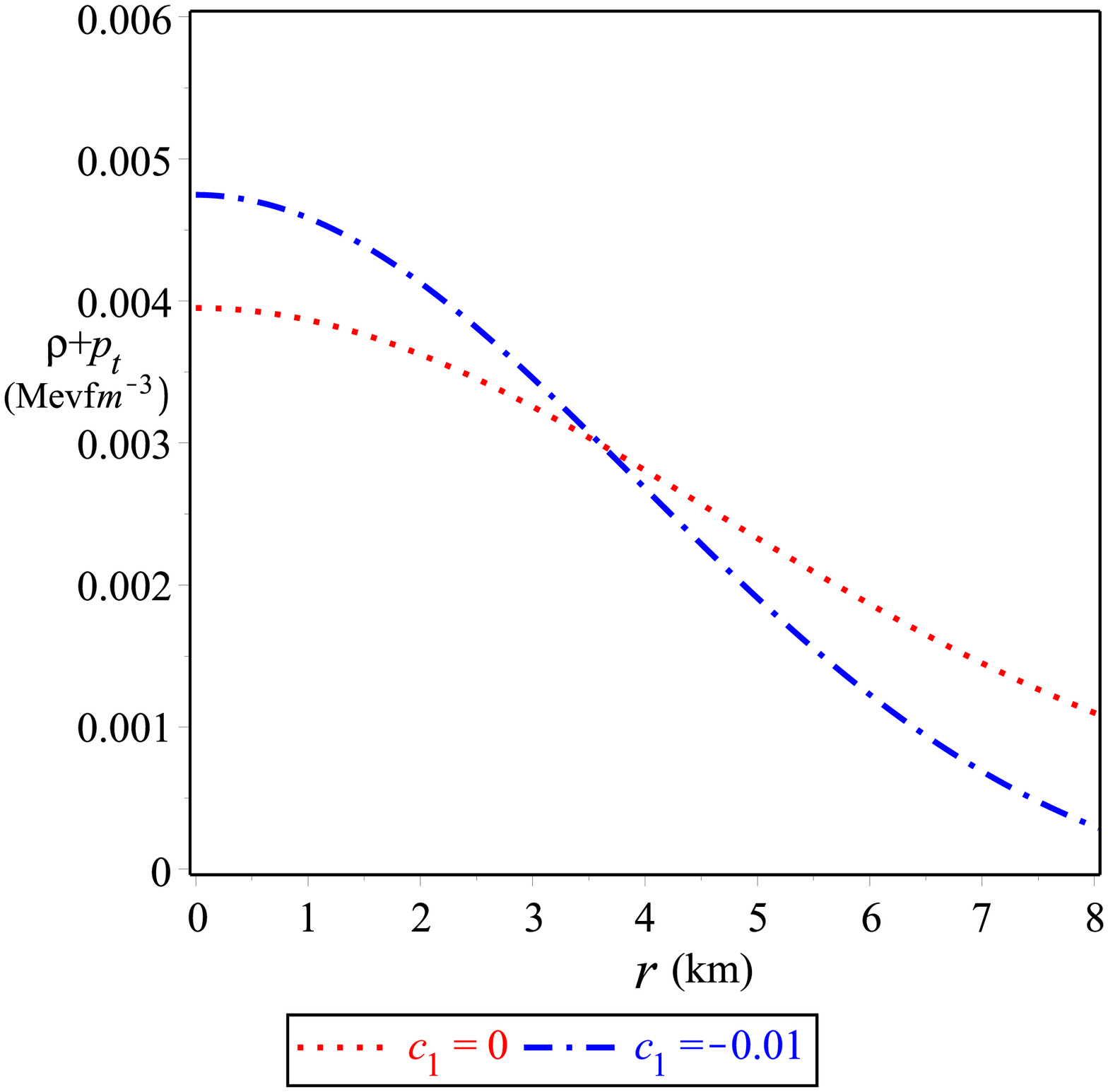}}
\subfigure[~ SEC, $(\rho-P_r-2P_\bot)$  (\ref{sol})]{\label{fig:EoS}\includegraphics[scale=.3]{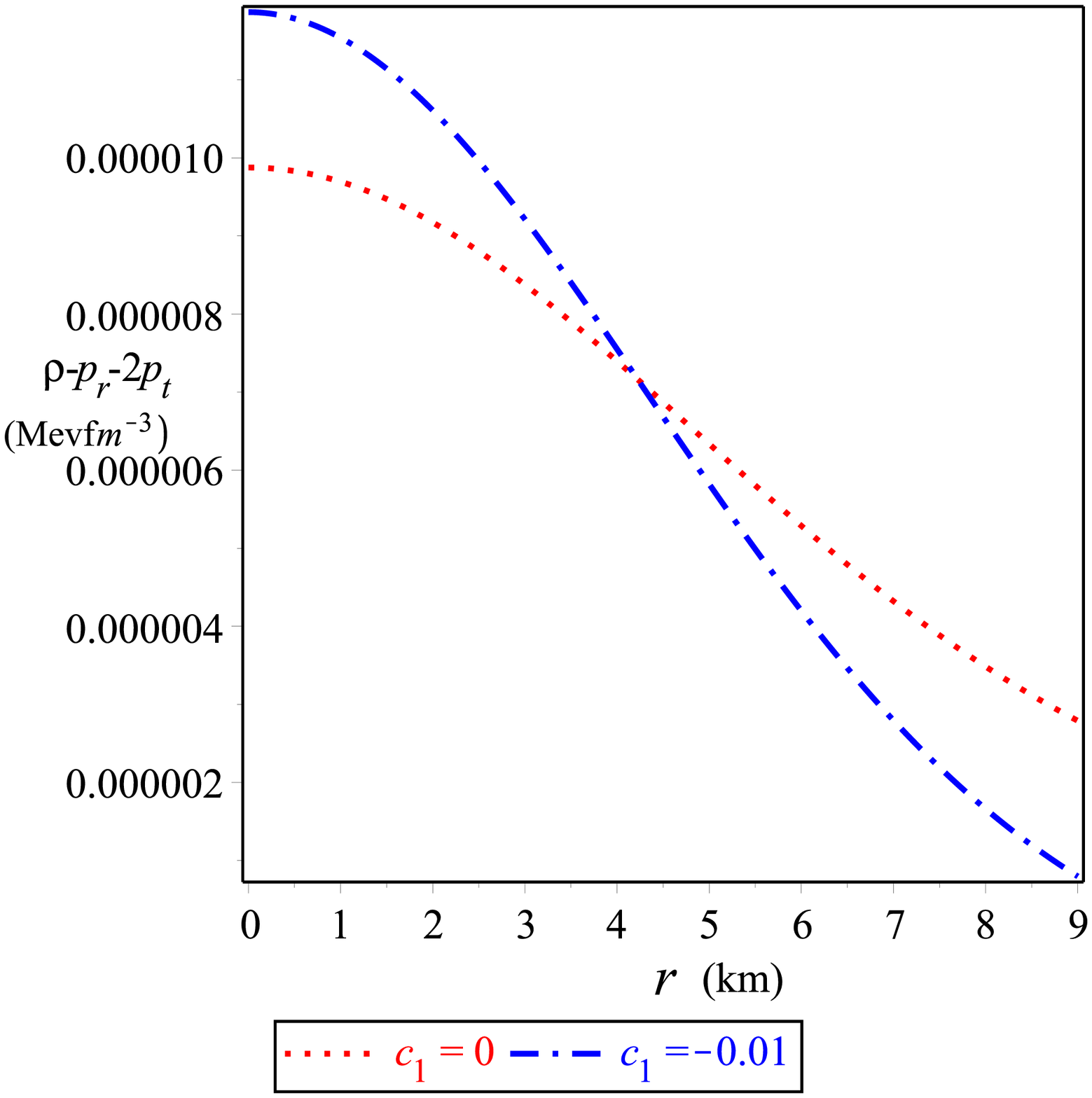}}
\caption[figtopcap]{\small{{WEC and SEC of solution  (\ref{sol}). Figures  \ref{Fig:6}, \ref{Fig:7}  and \ref{Fig:8} show that the energy conditions of our model are satisfied.}}}
\label{Fig:8}
\end{figure}

\subsection{\bf Mass-radius--relation}

For a spherically symmetric spacetime the compactification factor
$u(r)$ is defined as the ratio between its mass and radius. In the
compact stellar object, the compactification factor plays an
important role in understanding its physical properties. Using
solution (\ref{sol}),  the gravitational mass takes the following
form:
 \begin{eqnarray}\label{mass}
&& M(r)=4\pi{\int_0}^r  \rho \xi^2 d\xi=\frac{1}{32b_2{}^{7/2}\,r}\Bigg\{ 2e^{-b_0r^2}b_2{}^{3/2}\Bigg[\{4+4b_0c_1r^4+2r^2[b_0-3c_1]\}b_2{}^{2}-b_0b_2r^2[2b_0c_1r^2+9c_1+2b_0]\nonumber\\
 &&-3b_0{}^{2}c_1r^2\Bigg]+r\sqrt{\pi}b_2\,erf(\sqrt{b_2}r)[16b_2{}^3
 +10b_2{}^2(b_0-c_1)+b_0b_2(2b_0+9c_1)+3b_0{}^2c_1]+8b_2{}^{7/2}(c_1r^2-1)\Bigg\}\,,
\end{eqnarray}
where $erf(x)$ is the error function defined as follows:
\begin{eqnarray}
erf(x)=\frac{2}{\sqrt{\pi}}\int_0^x e^{-t^2}dt\,.\end{eqnarray} The compactification factor $u(r)$ is defined as follows:
\begin{eqnarray}
&&u(r)=\frac{M(r)}{r}=\frac{1}{32b_2{}^{7/2}\,r^2}\Bigg\{ 2e^{-b_0r^2}b_2{}^{3/2}\Bigg[\{4+4b_0c_1r^4+2r^2[b_0-3c_1]\}b_2{}^{2}-b_0b_2r^2[2b_0c_1r^2+9c_1+2b_0]\nonumber\\
 &&-3b_0{}^{2}c_1r^2\Bigg]+r\sqrt{\pi}b_2\,erf(\sqrt{b_2}r)[16b_2{}^3
 +10b_2{}^2(b_0-c_1)+b_0b_2(2b_0+9c_1)+3b_0{}^2c_1]+8b_2{}^{7/2}(c_1r^2-1)\Bigg\}\,.\end{eqnarray}
Fig. \ref{Fig:9} shows the behaviors of the gravitational mass and
compactification factor.
\begin{figure}
\centering
\subfigure[~Gravitational mass of  (\ref{sol}).]{\label{fig:mass}\includegraphics[scale=0.3]{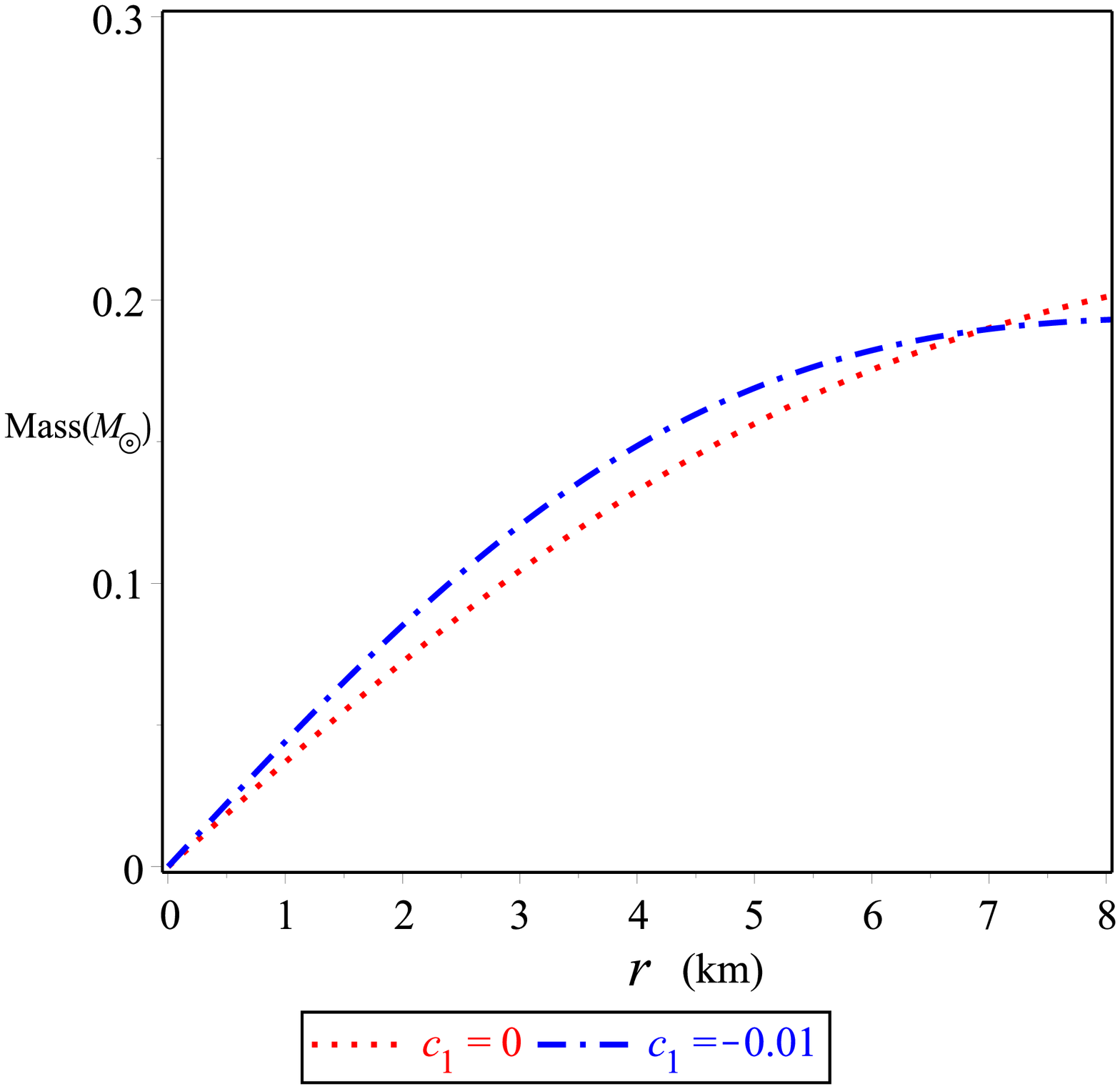}}
\subfigure[Compactification factor of  (\ref{sol}).]{\label{fig:u}\includegraphics[scale=.3]{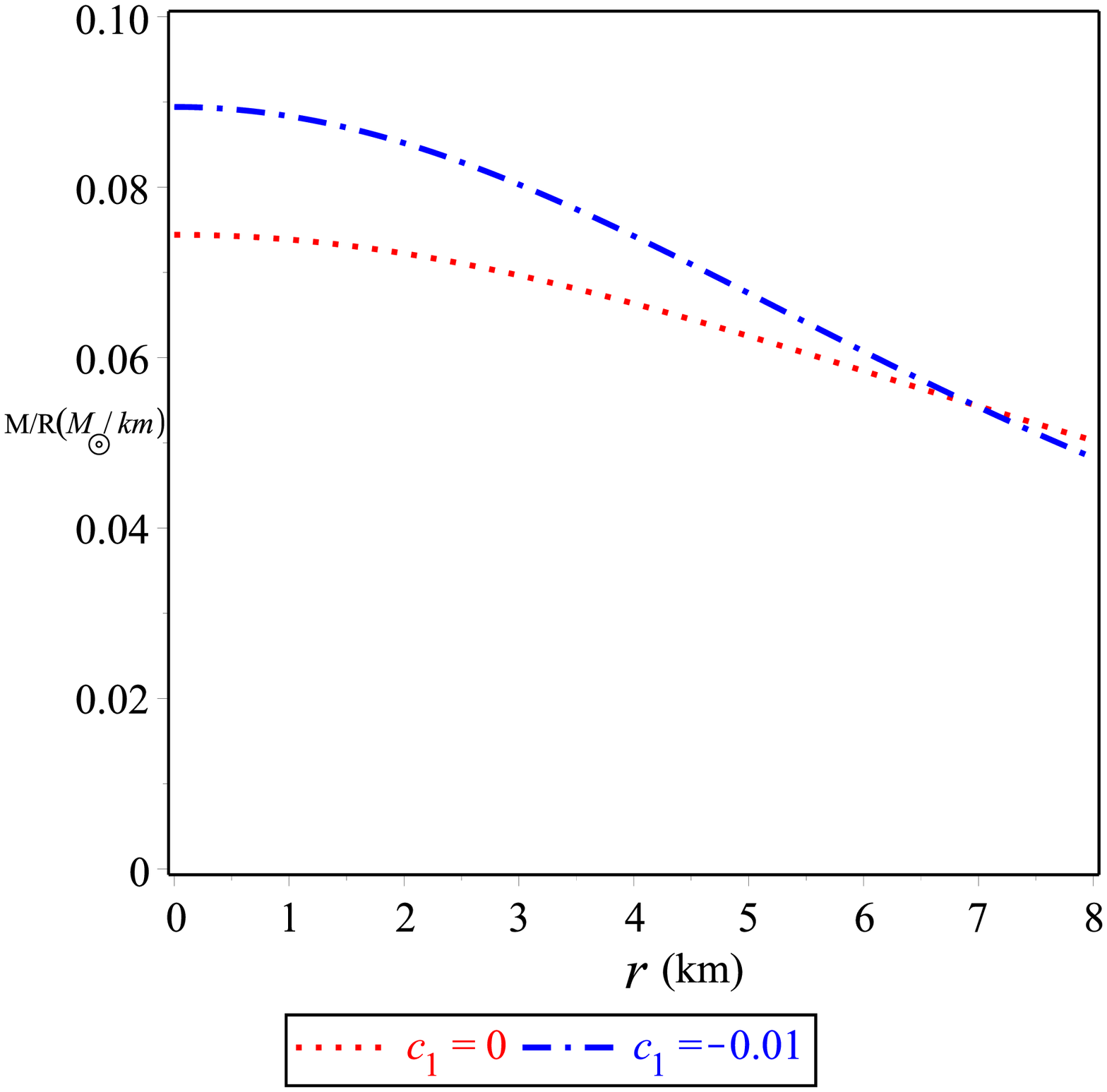}}
%
\caption[figtopcap]{\small{{Gravitational mass and compactification factor  of solution  (\ref{sol}).  Fig. \ref{Fig:9} \subref{fig:u} shows that at $c_1\neq 0$, the compactness factor has a  greater value than that at $c_1=0$. }}}
\label{Fig:9}
\end{figure}
As it is shown in Fig. \ref{Fig:9} \subref{fig:mass} the
gravitational mass increases as the radial coordinate increases,
contrary to Fig. \ref{Fig:9} \subref{fig:u} which reveals that the
compactification factor decreases as the radial coordinate
increases.

\subsection{\bf Equation of state (EoS) }

Das et al.  \cite{Das:2019dkn}  derived the  EoS  for a neutral
compact stellar object and showed that it is almost  linear;
however, in this study the EoS is nonlinear. This condition can be
explained by calculating the radial and transverse  EoS  that
respectively have the following form:
  \begin{eqnarray} \label{sol2}
 && \omega_r=\frac{P_r}{\rho}=\frac{e^{-b_2r^2}[1+b_0r^2+2r^4b_0c_1-r^4b_0{}^2-c_1r^6b_0{}^2+b_0b_2r^4+c_1b_0b_2r^6+2b_2r^2+5b_2c_1r^4-e^{b_2r^2}(1+c_1r^2)]}
 {e^{-b_0r^2}[b_0c_1(b_0-b_2)r^6+[a_0{}^2+(6c_1-b_2)b_0+3c_1b_2]r^4+(2b_2-4c_1+3b_0)r^2-1]+1+c_1r^2}\,, \nonumber\\
 &&\omega_\bot=\frac{P_\bot}{\rho}=\frac{1+c_1r^2-e^{-b_2r^2}[1+2r^2c_1-r^2b_0-2c_1r^4b_0+b_0b_2r^4+b_0b_2c_1r^6-b_0{}^2r^4-b_0{}^2c_1r^6+r^4c_1b_2]}
 {e^{-b_0r^2}[b_0c_1(b_0-b_2)r^6+[a_0{}^2+(6c_1-b_2)b_0+3c_1b_2]r^4+(2b_2-4c_1+3b_0)r^2-1]+1+c_1r^2}\,.
  \end{eqnarray}
  \begin{figure}
\centering
\subfigure[~Radial EoS ]{\label{fig:pr}\includegraphics[scale=0.3]{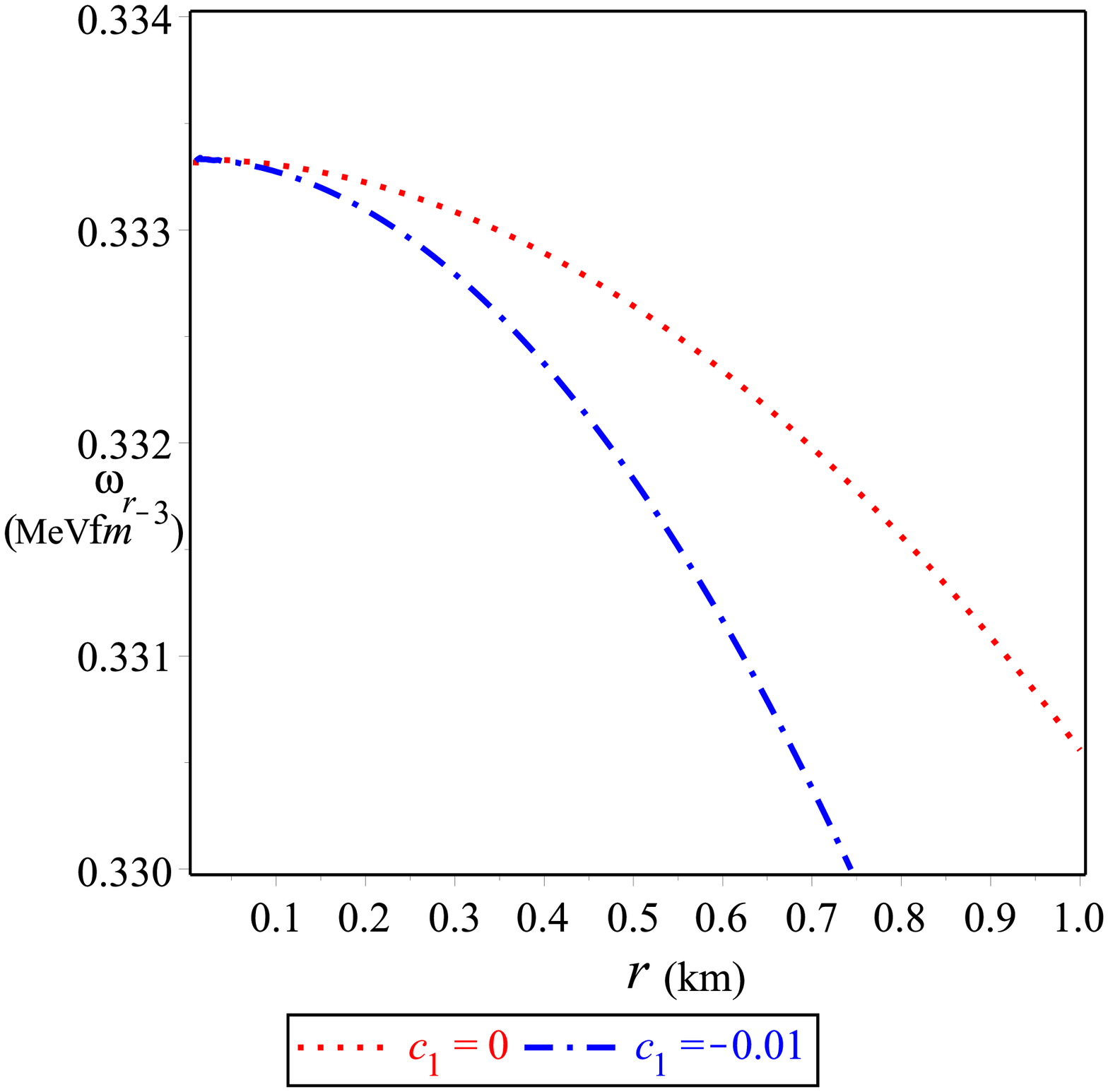}}
\subfigure[~Tangential EoS]{\label{fig:pt}\includegraphics[scale=.3]{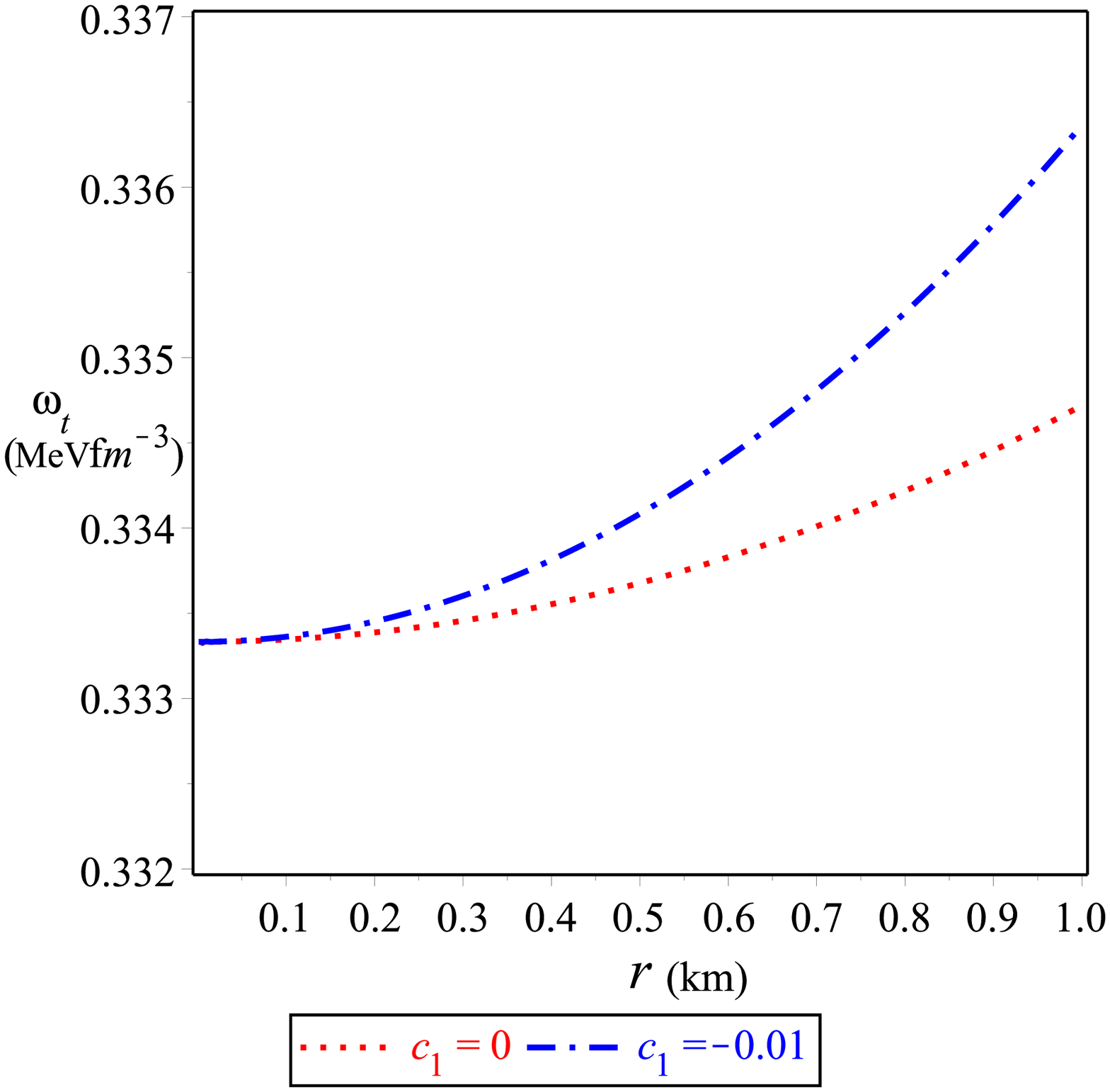}}
\caption[figtopcap]{\small{{Plots of the radial and transverse   EoS  of solution  (\ref{sol}. )}}}
\label{Fig:10}
\end{figure}
Fig. \ref{Fig:10} illustrates the behavior of the radial and
transverse EoS.  Figs. \ref{Fig:10} \subref{fig:pr}, and
\subref{fig:pt} show that the EoS is nonlinear. The  nature of the
metric potential given by Eq. (\ref{Eq3}) and perhaps indicates
the reason for the nonlinearity of the EoS. This phenomena can be
explained as follows, with the asymptotic forms of Eq.
(\ref{sol2}) assuming the form:
  \begin{eqnarray} \label{asym}
 \omega_r\simeq \frac{1}{3}-\frac{2b_0(4c_1-b_0)}{c_1-2b_0}r^2+O\Big(r^4\Big)\,, \qquad \omega_\bot\simeq \frac{1}{3}+\frac{b_0(4c_1-b_0)}{c_1-2b_0}r^2+O\Big(r^4\Big)\,,
  \end{eqnarray}
where  the constant $c_1$ has no effect on the nonlinearity of the
EoS. The effect of nonlinearity originates from the contribution
of the constant $b_0$.
\section{Stability of solution (\ref{sol})}\label{S6}
We now discuss the most critical condition which determines how
realistic is a compact stellar object, i.e., the stability
condtion. Here we investigate this  issue from the viewpoint of
the TOV equation and adiabatic index.
\subsection{Equilibrium Analysis Through TOV Equation}

In this subsection, we  discuss the stability of the derived
model. Accordingly, we assume a hydrostatic equilibrium governed
by the TOV equations. Using the TOV-equation
\cite{PhysRev.55.364,PhysRev.55.374,PoncedeLeon1993}, we obtain:
\begin{eqnarray}\label{TOV} \frac{2[P_\bot-P_r]}{r}-\frac{M_g(r)[\rho(r)+P_r]e^{[\alpha(r)-\beta(r)]/2}}{r}-\frac{dP_r}{r}=0,
 \end{eqnarray}
 where $M_g(r)$ is the gravitational mass confined in a
radius $r$ that is defined from the Tolman--Whittaker mass
formula using the following equation:
\begin{eqnarray}\label{ma} M_g(r)=4\pi{\int_0}^r\Big({T_t}^t-{T_r}^r-{T_\theta}^\theta-{T_\phi}^\phi\Big)r^2e^{[\alpha(r)+\beta(r)]/2}dr=\frac{r \alpha' e^{[\beta(r)-\alpha(r)]/2}}{2}\,,
 \end{eqnarray}
Using Eq. (\ref{ma}) in Eq. (\ref{TOV}), we obtain the following:
\begin{eqnarray}\label{ma1} \frac{2(P_\bot-P_r)}{r}-\frac{dP_r}{dr}-\frac{\alpha'[\rho(r)+P_r]}{2}=F_g+F_a+F_h=0\,,
 \end{eqnarray}
where $F_g=-\frac{\alpha'[\rho(r)+P_r]}{2}$,
$F_a=\frac{2(P_\bot-P_r)}{r}$ and $F_h=-\frac{dP_r}{dr}$ are the
gravitational,  anisotropic, and  hydrostatic forces respectively.
{ The solution of the TOV equation represented by model (\ref{sol})
is depicted in Fig. \ref{Fig:11}.}
 \begin{figure}
\centering
\subfigure[~TOV Equation when $c_1=0$ ]{\label{fig:TOV0}\includegraphics[scale=0.3]{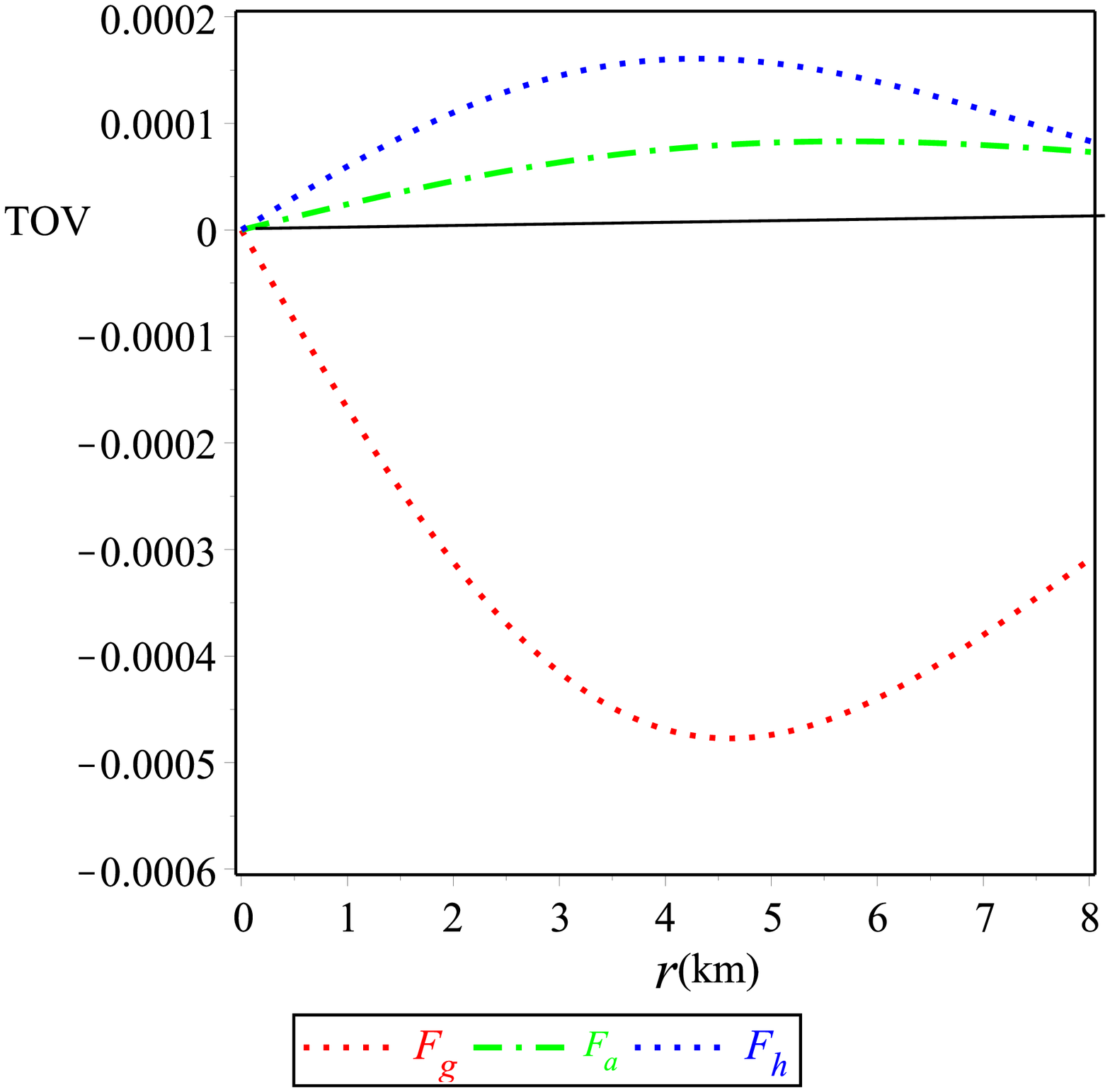}}
\subfigure[~ TOV Equation when $c_1=-0.01$]{\label{fig:TOV1}\includegraphics[scale=.3]{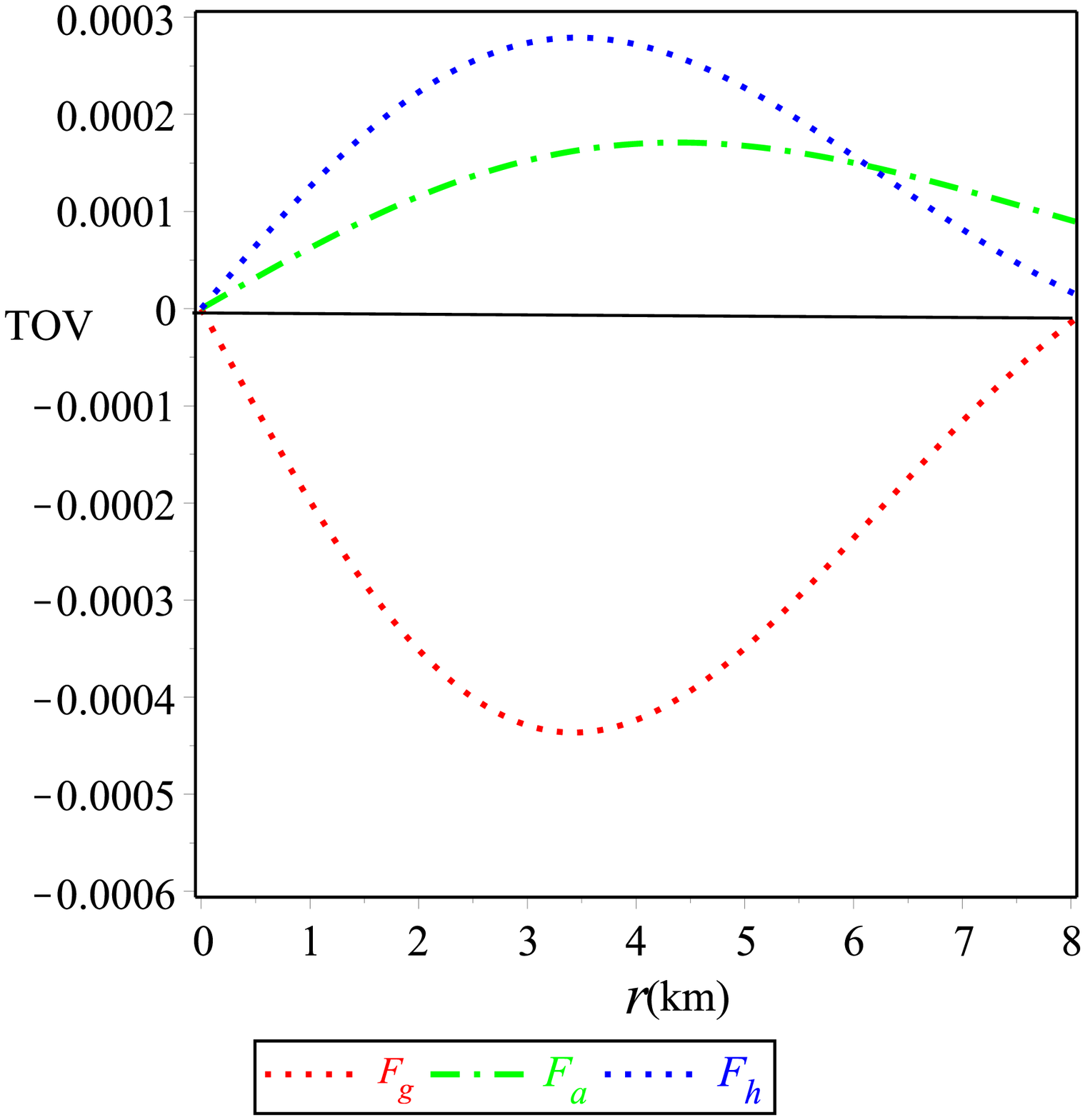}}
\caption[figtopcap]{\small{{TOV of solution  (\ref{sol}). }}}
\label{Fig:11}
\end{figure}
The three different forces are plotted in Fig. \ref{Fig:11}, which
shows that the hydrostatic and anisotropic forces are positive and
 dominated by the gravitational force which is negative, to
maintain the   hydrostatic equilibrium of the system.

Fig. \ref{Fig:11} \subref{fig:TOV1} shows that for $c_1=-0.01$,
the three different forces converge more rapidly than  for
$c_1=0$. Thus, in the non--vanishing $c_1$ case  of the
higher--order curvature case, the system tends to be more stable
than in the linear curvature case.

\subsection{\bf Adiabatic index}

The adiabatic index $\gamma$ is defined as, follows:
\begin{eqnarray}\label{ai} \gamma=\frac{\rho+P}{P}\frac{dP}{d\rho}\,.
 \end{eqnarray}
This index allows us to link the structure of a spherical
symmetric static object and the EoS of the interior solution, and
it helps in the study of the  stability of a stellar compact
object \cite{Moustakidis:2016ndw}. In order for the interior
solution to be stable, its adiabatic index must be greater than
$3/4$ \cite{1975A&A....38...51H} and when  $\gamma=\frac{4}{3}$,
the isotropic sphere will be in neutral equilibrium. According to
the work of Chan et al. \cite{10.1093/mnras/265.3.533}  the
condition $\gamma >\Gamma$  for the stability of a relativistic
anisotropic sphere should be satisfied,  where $\Gamma$ is
determined as follows: \begin{eqnarray}\label{ai}
\Gamma=\frac{4}{3}-\left\{\frac{4(P_r-P_\bot)}{3\lvert
P'_r\lvert}\right\}_{max}\,.
 \end{eqnarray}
Figure \ref{Fig:12} shows that the stability condition of model
(\ref{sol}) is verified according to the analysis of two adiabatic
indexes because both have values greater than $\frac{4}{3}$.  Fig.
\ref{Fig:12} \subref{fig:c=1} shows that the adiabatic index
$\gamma$ at $c_1=-0.01$ has a greater value  than that at $c_1=0$,
which means that the case that  differs from GR  is more stable
than the case of GR itself.

{ In table \ref{Table1}  we present different pulsars to calculate the two constants, $b_0$ and $b_1$, that characterized our model. In tables \ref{Table2} and \ref{Table2} we use the values of the constants $b_0$ and $b_1$ that are calculated in table \ref{Table1} to calculate the energy-density, radial and tangential velocities strong condition at the center and at the boundary of the pulsars presented in table \ref{Table1} for the GR and $\mathrm{f(R)}$.}
\begin{figure}
\centering
\subfigure[~Adiabatic index of  (\ref{sol}) when $c_1=0$]{\label{fig:c=0}\includegraphics[scale=0.3]{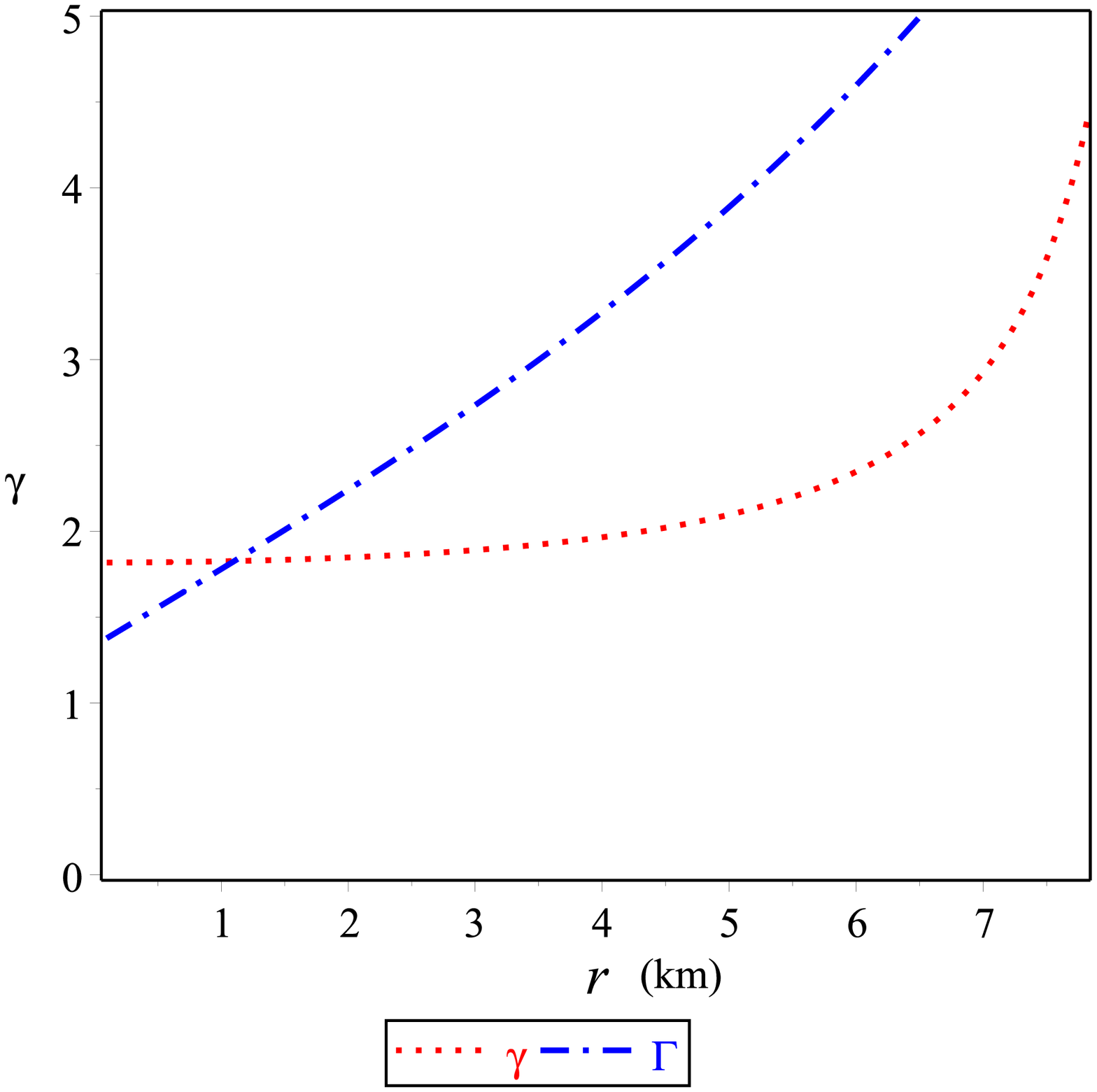}}
\subfigure[~Adiabatic index of  (\ref{sol}) when $c_1=-0.01$]{\label{fig:c=1}\includegraphics[scale=.3]{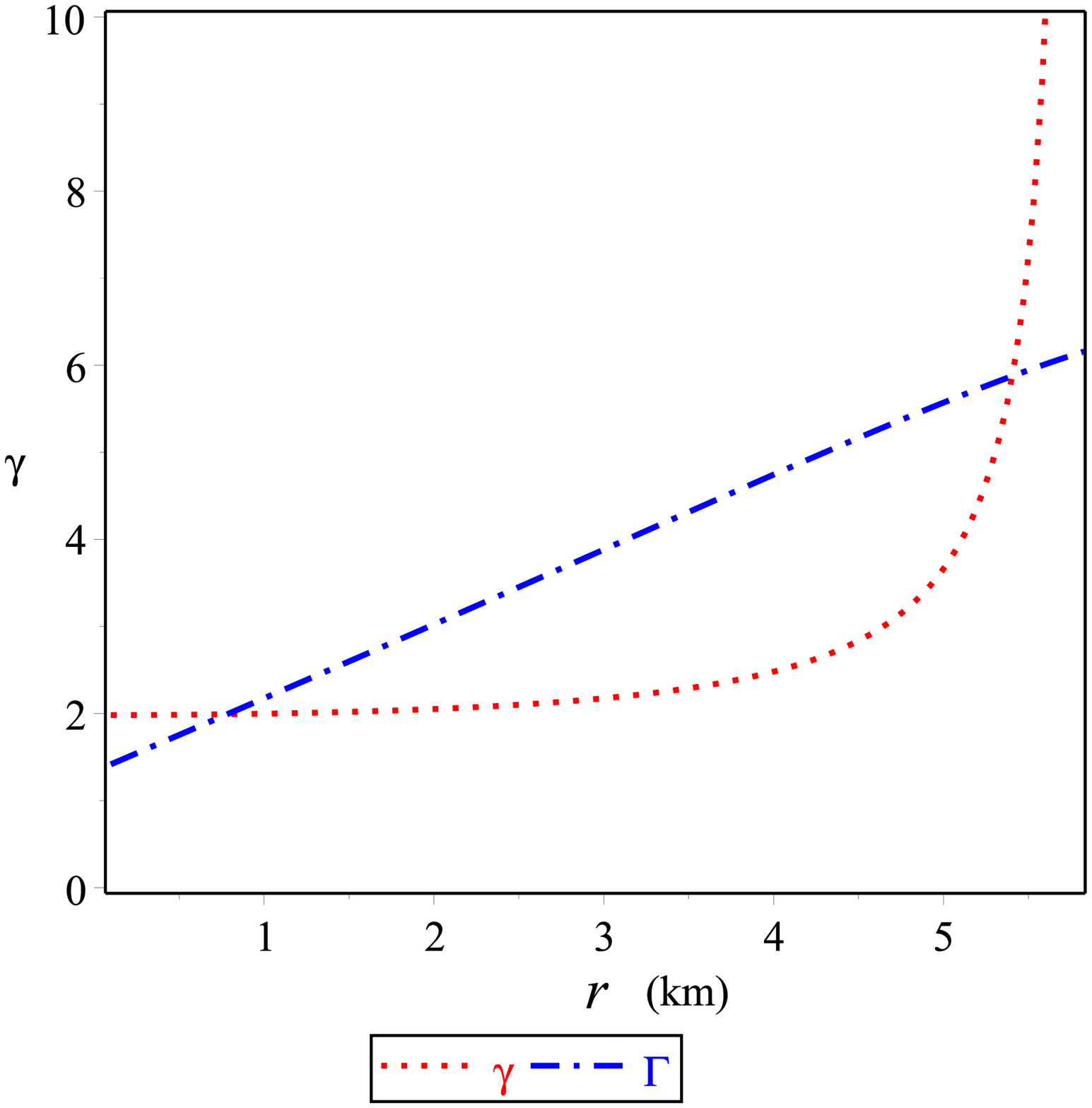}}
%
\caption[figtopcap]{\small{{Adiabatic index  of solution  (\ref{sol}).}}}
\label{Fig:12}
\end{figure}

\begin{table*}[t!]
\caption{\label{Table1}%
Values of model parameters}
\begin{ruledtabular}
\begin{tabular*}{\textwidth}{lcccccc}
{{Pulsar}} & Mass ($M_{\odot}$) & {Radius (km)} & {$b_0$} & {$b_1$} & \\ \hline
&&&&&&\\
EXO 1785 - 248 & $1.3\pm 0.2$ & $8.849\pm0.4$ & 0.03034302235 & -4.332109774& \\
&&&&&&\\
Cen X-3 & $1.49\pm 0.08$ & $9.178\pm0.13$ &0.03971381998& -6.502447317& \\
&&&&&&\\
RX J 1856 -37 & $0.9\pm 0.2$ & $\simeq 6$ & 0.05184675045 & -3.732966031 & \\
&&&&&&\\
4U1608 - 52 & $1.74\pm 0.14$ & $9.52\pm0.15$ & 0.04899708732 & -9.163327477 &\\
&&&&&&\\
Her X-1 & $0.85\pm 0.15$ & $8.1\pm0.41$ & 0.02481137166 & -2.934495513 &  \\
&&&&&&\\

\end{tabular*}
\end{ruledtabular}
\end{table*}
\newpage
\begin{table*}[t!]
\caption{\label{Table2}%
Values of physical quantities at $c_1=0$}
\begin{ruledtabular}
\begin{tabular*}{\textwidth}{lcccccccccc}
{{Pulsar}} &{$\rho\lvert_{_{_{0}}}$} & {$\rho\lvert_{_{_{b}}}$} & {$\frac{dP_r}{d\rho}\lvert_{_{_{0}}}$} & {$\frac{dP_r}{d\rho}\lvert_{_{_{b}}}$}
  & {$\frac{dP_\bot}{d\rho}\lvert_{_{_{0}}}$} & {$\frac{dP_\bot}{d\rho}\lvert_{_{_{b}}}$}&{$(\rho-P_r-2P_\bot)\lvert_{_{_{0}}}$}&{$(\rho-P_r-2P_\bot)\lvert_{_{_{b}}}$}&{$z\lvert_{_{_{b}}}$}&\\ \hline
&&&&&&&&&&\\
EXO 1785 - 248&0.3624$\times10^{-2}$&0.593$\times10^{-3}$ &-1 &0.2935 &-0.06667&0.3532&9.952$\times10^{-13}$ & 1.1$\times10^{-13}$&7.724\\
&&&&&&&&&&\\
Cen X-3 & 0.4743$\times10^{-2}$ &0.3868$\times10^{-3}$ &-1 & 0.1363 &-0.6 &0.4319&7.962$\times10^{-13}$&2$\times10^{-13}$&24.822 \\
&&&&&&&&&&\\
RX J 1856 -37 &0.61919$\times10^{-2}$&0.1266$\times10^{-2}$&0.6 &0.326&0.36&0.337&0&0&5.4656 \\
&&&&&&&&&&\\
4U1608 - 52 & 0.5851$\times10^{-2}$ &0.261$\times10^{-3}$ & -1 &-0.14778&0 &0.57389&0&0&96.68\\
&&&&&&&&&&\\
Her X-1 & 0.2963$\times10^{-2}$ &0.828$\times10^{-3}$ &-.3333 &0.3632 & 0.6667&0.318 &0&-2$\times 10^{-13}$& 3.337 \\
\end{tabular*}
\end{ruledtabular}
\end{table*}
\begin{table*}[t!]
\caption{\label{Table3}%
Values of physical quantities at $c_1=-0.01$}
\begin{ruledtabular}
\begin{tabular*}{\textwidth}{lcccccccccc}
{{Pulsar}} &{$\rho\lvert_{_{_{0}}}$} & {$\rho\lvert_{_{_{b}}}$} & {$\frac{dP_r}{d\rho}\lvert_{_{_{0}}}$} & {$\frac{dP_r}{d\rho}\lvert_{_{_{b}}}$}
  & {$\frac{dP_\bot}{d\rho}\lvert_{_{_{0}}}$} & {$\frac{dP_\bot}{d\rho}\lvert_{_{_{b}}}$}&{$(\rho-P_r-2P_\bot)\lvert_{_{_{0}}}$}&{$(\rho-P_r-2P_\bot)\lvert_{_{_{b}}}$}&{$z\lvert_{_{_{b}}}$}&\\ \hline
&&&&&&&&&&\\
EXO 1785 - 248&0.4221$\times10^{-2}$  & 0.4045$\times10^{-4}$  & -1 &-0.2163 & -0.6667$\times10^{-1}$ &0.6081&9.952$\times10^{-13}$ &7$\times10^{-13}$ & 7.724\\
&&&&&&&&&&\\
Cen X-3 & 0.534$\times10^{-2}$ &0 & -1 & -1.021&-0.6&1.0106&7.9618$\times10^{-13}$&1.1$\times10^{-13}$& 24.822 \\
&&&&&&&&&&\\
RX J 1856 -37 &0.6789$\times10^{-2}$&0.6724$\times10^{-3}$&0.6 &0.1503&.0.36&0.4248&0&-1$\times10^{-13}$&5.4656 \\
&&&&&&&&&&\\
4U1608 - 52 & 0.645$\times10^{-2}$ & 0 & -1 &-1.794&0 &1.3972&-7.962$\times10^{-13}$&-2$\times10^{-14}$&96.68\\
&&&&&&&&&&\\
Her X-1 & 0.356$\times10^{-2}$ & 0.2068$\times10^{-3}$ &-0.3334 &0.141 & 0.6667&0.429&0&-3$\times10^{-13}$& 3.337 \\
\end{tabular*}
\end{ruledtabular}
\end{table*}

\section{\bf Concluding remarks}\label{S7}

In this study, we studied compact stellar objects in
$\mathrm{f(R)}$ gravity. We have applied the non-vacuum field
equations of $\mathrm{f(R)}$ after rewriting them in terms of
$f_R=\mathrm{\frac{df(R)}{dR}}$ to a spherically symmetric
spacetime. We obtained a system of four nonlinear differential
equations comprising  six unknown functions, the three components
of the energy--momentum tensor, ($\rho(r)$, $P_r$, $P_\bot$), the
two components of the metric potentials, and the form
$f_R=\mathrm{\frac{df(R)}{dR}}$. To solve such a system we have
assumed the form of the metric potentials, given by Krori-Barua
ansatz  that contain three constants. As a result the system was
rendered easy to be solved analytically. We have derived the three
components of the energy--momentum and the form of
$\mathrm{f(R)}$. We have shown that the form of the Ricci scalar
associated with this compact star is not trivial and  the
asymptotic form of $\mathrm{f(R)}$ behaves as a polynomial
function. This solution contains four constants of integration.
One of these constants caused the deviation of our solution from
the GR models, leading to the higher--order curvature terms. When
this constant was set equal to zero, we recovered the GR compact
star solution. In order to further simplify the system, we have
assumed two constants of the metric potential to be equal and have
applied the matching condition to the metric derived in
\cite{Nashed:2019tuk}, which has a nontrivial form of the Ricci
scalar; the metric is also different from a Schwarzschild one and
determines the relation between two constants and the mass and
radius of a compact star, leaving the constant responsible for the
deviation from GR to be arbitrary.

We have listed the necessary conditions that any non-vacuum
solution must satisfy in order to become compatible with a real
compact star. We have shown that the three components of the
energy--momentum tensor satisfied the listed conditions for a real
star. Moreover, we have studied  the energy conditions, namely,
the WEC, DEC and SEC and have shown that the present solution
satisfied all of these conditions. In addition, we have investigated the stability of the
derived solution by calculating  adiabatic index and showed
that it is greater than $4/3$ as required  \cite{10.1093/mnras/265.3.533}. { It is interesting to discuss our solutions in the context of more compact objects like neutron stars, to make contact with events like the $GW190814$. These solutions have been studied in \cite{Astashenok:2021peo} in the context
of $\mathrm{f(R)}$ gravity. In our case, extra caution is needed since our approach applies to inhomogeneous solutions.
Nevertheless, pulsars with spin less than $3ms$ or even the product of the merging of two neutron stars if it is a neutron star, can
initially be quite inhomogeneous, so during the ring-down, our solution could be relevant. We hope to address this issue in future work since such a study would require the implementation of a numerical recipe appropriately tailored to our solutions.}

 In conclusion, we have succeeded for the first time to derive a
nontrivial  anisotropic compact star  in   $\mathrm{f(R)}$ by
assuming a specific form for the metric potential. This study can
be continued by searching for a constraint other than the form of
the metric potential to achieve a closed form of the system of
field equations of   $\mathrm{f(R)}$, like to assume a specific
form of the EoS. We expect that the physics of the resulting model
will be entirely different  from that presented in this study. We
hope to address this issue in the future.
  \begin{center}
{\bf{Appendix A}}
\end{center}
\begin{center}
{{\b The form of $f(r)$ }}
\end{center}
Using the fact that $F=\frac{df(R)}{dR}=\frac{df(r)}{dr}\frac{dR(r)}{dr}$, we can get the form of $f(r)$ by solving the
following differential equation:
   \begin{eqnarray}\label{f(r)}
   &&8f-\frac{1}{8(1-(1+b_0b_2r^4+r^2[b_0-2b_2])(1+r^2[b_0-b_2])e^{-b_2r^2})^{^{^3}}} \Bigg\{rf'\Bigg[e^{-3b_2r^2}\Bigg(4b_0{}^2b_2{}^2r^{16}[b_0-b_2]^2(b_0{}^2-4b_0b_2-3b_2{}^2)-68\nonumber\\
 &&
   +2b_0b_2[24b_2{}^4-21b_0b_2{}^3-70b_0{}^2b_2{}^2+47b_0{}^3b_2-4b_0{}^2][b_0-b_2]r^{14}+4(b_0{}^6-12b_2{}^6-24b_0{}^5b_2-158b_0{}^3b_2{}^3+30b_0b_2{}^5
  \nonumber\\
 && +107b_0{}^4b_2{}^2+50b_0{}^2b_2{}^4)r^{12}+2(9b_0{}^5-28b_2{}^5-95b_0{}^4b_2+286b_0b_2{}^4+364b_0{}^3b_2{}^2-564b_0{}^2b_2{}^3)r^{10}
   -4r^8(b_0{}^4+144b_0b_2{}^3\nonumber\\
 &&-8b_0{}^3b_2-57b_2{}^4-47b_0{}^2b_2{}^2)-2r^6(351b_0b_2{}^2-234b_0{}^2b_2+6b_0{}^3-37a_2{}^3)-4(109b_2{}^2-124b_0b_2+31b_0{}^2)r^4
   -6(21b_2+b_0)r^2\Bigg)\nonumber\\
 &&+4e^{-2b_2r^2}\Bigg(33-b_0{}^2b_2{}^2(b_0-b_2)^2r^{12}-b_0b_2{}^2(3b_0+5b_2)(b_0-b_2)r^{10}-4(11b_0{}^3b_2-84b_0{}^2b_2{}^2-14b_2{}^4-
   b_0{}^4+95b_0b_2{}^3)r^8\nonumber\\
 &&+4(3b_0{}^3-109b_2{}^3-149b_0{}^2b_2+296b_0b_2{}^2)r^6+4(32b_0{}^2-131b_0b_2+95b_2{}^2)r^4+12(b_0+9b_2)r^2\Bigg)-1-2e^{-b_2r^2}
   \Bigg(30\nonumber\\
 &&-4b_0b_2(b_0-b_2)r^6-2(b_0{}^2+4b_2{}^2-7b_0b_2)r^4-3(3b_2-b_2)r^2\Bigg)\Bigg]-6r^2f''\Bigg[e^{-3b_2r^2}\Bigg([1+b_0b_2r^4+(2b_2-b_0)r^2]
 \nonumber\\
 &&[1+(b_0-b_2)r^2][8+b_0b_2(b_0+3b_2)(b_0-b_2)r^8-(b_0{}^3+3b_0{}^2b_2-14b_0b_2{}^2+6b_2{}^3)r^6+b_2(b_0+3b_2)r^4
 +(b_0+7b_2)r^2]\Bigg)\nonumber\\
 &&-e^{-2b_2r^2}\Bigg(16-2b_0b_2(b_2{}^2-b_0{}^2)r^8+(10b_0{}^2b_2-2b_0{}^3-4b_2{}^2)r^6-(14b_2{}^2-8b_0{}^2+34b_0b_2)r^4+
 2(b_0+7b_2)r^2\Bigg)\nonumber\\
 &&+e^{-b_2r^2}\{8+r^2(b_0-b_2)\}\Bigg]-6r^3f'''\Bigg[(1+b_0b_2r^4+[2b_2-b_0]r^2)^2(1+[b_0-b_2]r^2)^2e^{-3b_2r^2}\nonumber\\
 &&
 -e^{-2b_2r^2}(2-2b_0b_2[b_2-b_0]r^6+2[b_0{}^2-4b_0b_2+2b_2{}^2]r^4-2b_2r^2)+e^{-b_2r^2}\Bigg]\Bigg\}=0\,.
 \end{eqnarray}
 The above differential equation  is not easy  to  analytically solve therefore, we are shall find some approximate asymptotic solutions. Asymptotically and by putting
  $b_2=b_0$ we get,
 \begin{eqnarray} \label{fa(r)}
 3f'''+16b_0f''+8b_0f'-4b_0{}^2f=0
 %
    \end{eqnarray}
The solution of the above differential equation is  lengthy and here we write its asymptotic that takes the following form,
    \begin{eqnarray}\label{fas(r)}
    f(r)=c_2+c_3\sin\left(\frac{\sqrt{24b_0}r}{3}\right)+c_4\cos\left(\frac{\sqrt{24b_0}r}{3}\right)\,.
     \end{eqnarray}
In order for the solution (\ref{fas(r)}) to be compatible with the
form of $F(r)=\frac{df(R)}{dR}=\frac{df(r)}{dr}\frac{dr}{dR}$ up
to leading order,  we must assume $c_3=0$, and
$c_4=-\frac{3b_0}{4}=\frac{27c_1}{16}$ which results in
$b_0=\frac{9c_1}{4}$. Eq. (\ref{sol}) shows that when $c_1=0$ we
return to the case of GR because $F=1$, which results in $f(R)=R$.
Thus, the terms that contain $c_1$ make the solution (\ref{sol})
different from GR. Henceforth, we assume $b_2=b_0$ to make the
calculations more easy to handle. Using Eq. (\ref{mets}) and
constraints $b_2=b_0$, we obtain the Ricci scalar up at leading
order, which has the following form,
   \begin{equation} \label{Ris}
   R\approx b_0{}^2 r^2\, , \quad {\textrm which \, \, \, \, \, leards\, \, \, \, to\, \, \, \, \, } r=\pm\frac{\sqrt{R}}{b_0}.
   \end{equation}
   Using Eq. (\ref{Ris}) in Eq. (\ref{fas(r)}) we obtain the following form of $\mathrm{ f(R)}$:
    \begin{equation} \label{fa(R)}
    \mathrm{ f(R)}=c_2+c_4\cos\Big(\frac{\sqrt{24R}}{3\sqrt{b_0}}\Big)\approx c_2+c_4+R-\frac{c_1}{4b_0{}^2}R^2+\frac{c_1}{45b_0{}^3}R^3-\cdots \cdots\,.
    \end{equation}
      Equation  (\ref{fa(R)})  represents GR plus higher-order corrections so with corresponding choices of parameters such theory easily pass cosmological and astrophysical tests being realistic theory.
    \subsection*{Acknowledgments}
 This work was supported by MINECO (Spain), project PID2019-104397GB-I00 and PHAROS COST Action (CA16214) (SDO).
%

\end{document}